\documentclass{article}
\usepackage[top=0.85in,left=1.in,right=1.in,footskip=0.75in]{geometry}

\usepackage{amsmath,amssymb}
\usepackage{cite}

\usepackage{caption} 
\usepackage{color}
\usepackage{authblk}
\usepackage{enumitem}
\usepackage{microtype}
\usepackage{graphicx, amsmath}
\usepackage{booktabs} 
\usepackage{amsfonts} 
\usepackage{amssymb} 
\usepackage{algorithm}
\usepackage[noend]{algpseudocode}
\usepackage{float} 
\usepackage{hyperref}
\usepackage{url}
\usepackage{csquotes}

\usepackage{rotating}
\usepackage{url}
\usepackage{lineno}

\providecommand{\keywords}[1]
{
  \small	
  \textbf{\textit{Keywords---}} #1
}

\providecommand{\runningtitle}[1]
{
  \small	
  \textbf{\textit{Running title---}} #1
}

\title{The scaling of goals via homeostasis: an evolutionary simulation, experiment and analysis}

\author[1]{Léo Pio-Lopez}
\author[2]{Johanna Bischof}
\author[3]{Jennifer V. LaPalme}
\author[1,4*]{Michael Levin}

\affil[1]{Allen Discovery Center, Tufts University, Medford, MA, USA}
\affil[2]{Euro-BioImaging ERIC Bio-Hub, EMBL, Germany}
\affil[3]{University of Massachusetts Medical School, MA, USA}
\affil[4]{Wyss Institute for Biologically Inspired Engineering, Harvard University, Boston, MA 02115, USA}
\affil[*]{corresponding author: michael.levin@tufts.edu}

\date{}

\begin{document}

\maketitle
\begin{abstract}
All cognitive agents are composite beings. Specifically, complex living agents consist of cells, which are themselves competent sub-agents navigating physiological and metabolic spaces. Behavior science, evolutionary developmental biology, and the field of machine intelligence all seek an answer to the scaling of biological cognition: what evolutionary dynamics enable individual cells to integrate their activities to result in the emergence of a novel, higher-level intelligence that has goals and competencies that belong to it and not to its parts? Here, we report the results of simulations based on the TAME framework, which proposes that evolution pivoted the collective intelligence of cells during morphogenesis of the body into traditional behavioral intelligence by scaling up the goal states at the center of homeostatic processes. We tested the hypothesis that a minimal evolutionary framework is sufficient for small, low-level setpoints of metabolic homeostasis in cells to scale up into collectives (tissues) which solve a problem in morphospace: the organization of a body-wide positional information axis (the classic French Flag problem). We found that these emergent morphogenetic agents exhibit a number of predicted features, including the use of stress propagation dynamics to achieve its target morphology as well as the ability to recover from perturbation (robustness) and long-term stability (even though neither of these was directly selected for). Moreover, we observed unexpected behavior of sudden remodeling long after the system stabilizes. We tested this prediction in a biological system – regenerating planaria – and observed a very similar phenomenon. We propose that this system is a first step toward a quantitative understanding of how evolution scales minimal goal-directed behavior (homeostatic loops) into higher-level problem-solving agents in morphogenetic and other spaces.
\end{abstract}
\vspace{0.5cm}
\runningtitle{From metabolism to form}\\

\section{Introduction}

How does the complex anatomical structure of the body form reliably from the activity of individual cells? More than emergent morphogenesis, many organisms are able to reach the same specific target morphology despite surgical injury, changes of cell size or number, organ rearrangement, genetic defects, and many other insults \cite{pezzulo2016top}. How does the collective decision-making of cell groups enable regeneration, regulative, development, metamorphosis, and cancer suppression? Answers to these questions are critical to many fields, including evolutionary developmental biology, cell biology, and complexity science \cite{kauffman1993origins, roli2020emergence}. Beyond their importance for basic science, massive implications for transformative biomedicine rest on the ability to control the collective behavior of cells toward specific anatomical shapes. How can we best control morphogenetic systems composed of several levels of organization from cell to tissue, to achieve desired outcomes in regenerative medicine and synthetic bioengineering \cite{pezzulo2015re}? 

All of these questions raise the problem of understanding how the competencies of cells relate to those of the emergent tissue-level swarm, and what evolutionary dynamics enable this to occur. No individual cell knows what a finger is or how many a salamander is supposed to have, and yet when amputated, the collective will actively build until the precisely right number, size, shape, and position of the fingers is restored \cite{birnbaum2008slicing}. What is the origin of the dynamics that enable a cellular collective to work toward large-scale anatomical target states which are too complex and big for individual cells to represent?

Some frameworks, motivated by cybernetics and behavior science, seek a deep symmetry among homeostatic behavior across scales and material implementations \cite{fields2020scale, levin2021technological}. It has been suggested \cite{e24060819} that evolution pivoted some of the same strategies for navigating problem spaces across metabolic, anatomical, and behavioral domains. For example (reviewed in \cite{levin2021bioelectric, fields2020morphological}), the same mechanisms using ion channels, electrical synapses, and computations via bioelectrical networks underlie problem-solving in physiological space by bacterial communities \cite{prindle2015ion, yang2020encoding}, navigation of morphospace by a wide range of embryonic and regenerative systems \cite{levin2018bioelectric, harris2021bioelectric, bates2015ion}, and classical behavior in the 3D world by organisms with neurons and brains \cite{friston2010action, pezzulo2015active}.

Traditional behavior science focuses on the cognitive capacities of an "individual": memory, perception, learning, anticipation, decision-making, goal-directed behaviour and make two assumptions. First, the body is usually conceived as a fixed structure and thus follows the mainstream paradigm that the genome codes for specific bodyplans. Second, brain structure is thought of as stable - the individuality of its neuronal cells is gone for good. Our primary experience is that of a centralized, coherent Self. For organisms like caterpillar, which undergo metamorphosis, with drastic changes of body, brain, and behavior, studies usually focus on their separate phases of life. The transitional states are usually not studied \cite{levin2019computational}. 

However, all intelligences are collective intelligences. We are all made of parts that are themselves biological agents.  Regenerative biology and the development of new organisms like biobots \cite{kamm2018perspective} reveal that studies of cognition in intact bodies present just a narrow slice of a much bigger picture: that of the multi-scale interface between body and mind. Indeed, cognition is tightly linked to the physical structure of the body, from the anatomical level to the molecular and bioelectric levels that store its information \cite{levin2019computational}. And structure and function are both highly plastic. Cognition continues to function despite important changes to the body/brain and the corresponding modification of its information at the cellular, molecular, or bioelectric level \cite{levin2020life}. 

In the age of regenerative medicine, and of body and mind extension via brain-machine interfaces or neural implants, it seems essential to understand how mind and body co-evolve \cite{tarnita2013evolutionary}, in order to explore the origin of multicellularity and the scaling of basal cognition of individual cells into larger organisms. Understanding the \textquote{software} or algorithms that allow cells to perform goal-directed decision-making and the transition to larger selves is still poorly understood \cite{mathews2018body, levin2019computational}. In order to integrate and advance the fields of developmental/evolutionary biology, synthetic bioengineering, machine learning, and cognitive science, it is fundamental to begin to develop computational frameworks for asking how cognitive capacities arise in agents made of parts, and how these emergent subjects of cognition scale up during evolution. 

In the scale-free cognition framework \cite{levin2019computational}, we rely on ideas from control and information theory to identify multi-scale information processing principles. Earlier studies showed that individuals  can be outperformed by collectives \cite{woolley2010evidence}, and that their overall performance is dependent on several characteristics including their organisational or network structures \cite{mason2012collaborative}, the information aggregation and communication system among their individuals, and the diversity between their members \cite{hong2004groups}. These studies  focused mainly on human networks and the associated wisdom of the crowd. We focus here on morphogenetic systems.

Indeed, the mechanistic and algorithmic invariance between scales of organization has enabled numerous tools and approaches from neuroscience to be ported to developmental biology \cite{pezzulo2015re, pezzulo2021bistability, adams2014optogenetics} resulting in novel capabilities in regeneration, cancer reprogramming, and repair of birth defects. To move this field forward, and to better understand the role of evolution, it is essential to develop quantitative, generative models that reveal what dynamics are sufficient for the scaling of competencies and their generalization to new problem spaces. Thus, we sought to create a minimal in silico system in which we could observe whether, and how, evolutionary dynamics could make the shift from cell-level metabolic goals to the much larger, body-wide goal of patterning a positional information axis. Here, we present the details and analysis of a simple artificial embryogeny system in which we provide cell-level physiological homeostasis and observe the emergence of competence to reach a target morphology (solving the classic French Flag problem \cite{sharpe2019wolpert}) in an entirely new problem domain: anatomical morphospace. Our data show that evolutionary forces drive the emergence of several higher-level competencies, including error-minimization to reach an anatomical goal state and robustness to perturbation. This multi-scale homeostasis is driven by shared (non-local) stress dynamics and biologically-plausible cell-cell interactions. Our analysis reveals how minimal low-level metabolic requirements can scale up into morphogenetic competency.

\section{Foundation for the model}

Our model is based on the following background assumptions. 
The first is that individual cells already have the competency to stay alive via homeostatic pursuit of metabolic resources. This is a minimal requirement for all living forms (already present in microbes, long before multicellularity), but we assume its presence (i.e., we are not here  modeling the origin of life or the initial formation of the very first reproductive unit).

Second, we model two scales: a developmental phase where cells are alive and interact with each other, and an evolutionary wrapper which changes the frequency of different behavioral policies of the cells based on the fitness of the collective (how well the resulting “embryo” matches a very simple criterion: having a single, 3-valued positional information axis).

Third, our use of the term “goal” is not meant in any kind of meta-cognitive, self-aware complex sense. Goals here are synonymous with the cybernetic approach \cite{rosenblueth1943behavior, blac060}, in which certain systems can expend effort to reduce error from a specified homeostatic setpoint. Our aim is to show how such goals can scale in size and arise spontaneously in new problem spaces under evolutionary selection, given very minimal and realistic assumptions about the properties of the components.

Finally, while we do not explicitly invoke bioelectric signaling mechanisms, our cell interaction model is quite compatible with that mode of communication (spelled out in detail in \cite{levin2019computational}). We do provide the cells with gap junctions (electrochemical synapses \cite{mathews2017gap, miller2017electrical}), through which they can share state signals. We also include a mechanism for stress signals (initially, intracellular detectors of out-of-homeostasis states) to be propagated outward from stressed cells, as predicted by the TAME framework \cite{levin2021technological}.

\section{Related work: in silico embryogeny}

The \textquote{software} or set of of built-in behavioral and signaling policies  allowing cells to  cooperate and compete to reliably construct complex body pattern is still poorly understood \cite{mathews2018body, levin2019computational}. One relevant approach is amorphous computing, which refers to systems of many identical simple processors that have limited computational power and that interact locally. Typically, such systems have more a irregular structure in contrast to the regular structure of cell networks as they are used in our study \cite{abelson2000amorphous}. Several authors have investigated pattern formation in developmental biology using the same task we use in this article, the French flag problem. Herman and Liu  solved it through the simulation of linear iterative
arrays of cells \cite{herman1973daughter}. Several other models have been developed to resolve the same problem, including a reaction-diffusion model, use of computational primitives and genetic programming \cite{othmer1980scale, signes2016computational, miller2003evolving}. Miller evolved a development program that follows a predefined grammar that connects developmental encodings to genetic programming. He used a feed-forward Boolean circuit to implement a cell program  \cite{miller2003evolving}. Chavoya and Duthen  used a Genetic Algorithm to evolve Cellular Automata that produced different 2D and 3D shapes \cite{chavoya2006using} and evolved an Artificial Regulatory Network (ARN) for cell pattern generation, resolving the French flag problem \cite{chavoya2007use}. The combination of neural networks and cellular automata which grow under evolutionary control is described in the work of Elmenreich and Mordvintsev \cite{elmenreich2011evolving, mordvintsev2020growing}. A key difference in our work is that the artificial neural network (ANN) inside each cell here models the behavior of a gene-regulatory network and  can control gap junctions to deliver the morphogen and the ability to reduce stress (homeostasis) to the other cells.

Past studies have also been developed to investigate the multi-scale interface of body and mind, notably with  \textquote{morphological computation} in Artificial Life and Soft Evolutionary Robotics \cite{cheney2014evolved, pfeifer2006body, muller2017morphological, pfeifer2014cognition, bongard2011morphological}. These studies model and exploit the fact that brains, like other developing organs, are not hardwired but are able to ascertain the structure of the body and adjust their functional programs accordingly. Similarly to scale-free cognition, morphological computation is about connecting the body and cognition – showing how optimal control policies span different scales, mechanisms, and problem spaces. However, to our knowledge, no prior model has explicitly investigated how cell-level metabolic competencies scale up into tissue-level morphological ones.


\section{Methods}

\subsection{An in silico evolutionary system for the study of transitions from single-cell homeostasis to anatomical homeostasis of emergent tissue-level axial patterning}

\subsubsection{The general scheme of the evolutionary simulation}

\noindent\textbf{General system.}
We developed an evolutionary simulation system composed of cells that forms a tissue on a 2D grid. The system is an evolutionary, agent-based, spatialized model that uses two main time-scale loops. The outside loop is an evolutionary (phylogenetic) long time-scale where genomes are mutated and agents are selected. The inner loop is a short time-scale ontogenetic loop that constructs each agent from its genome, simulates the development, and then tests the phenotype for fitness. Each agent is one cell, which has a genome that encodes some very simple metabolic processes (interacts with the other cells). 

\begin{figure}[!pt]
  \includegraphics[width=\linewidth]{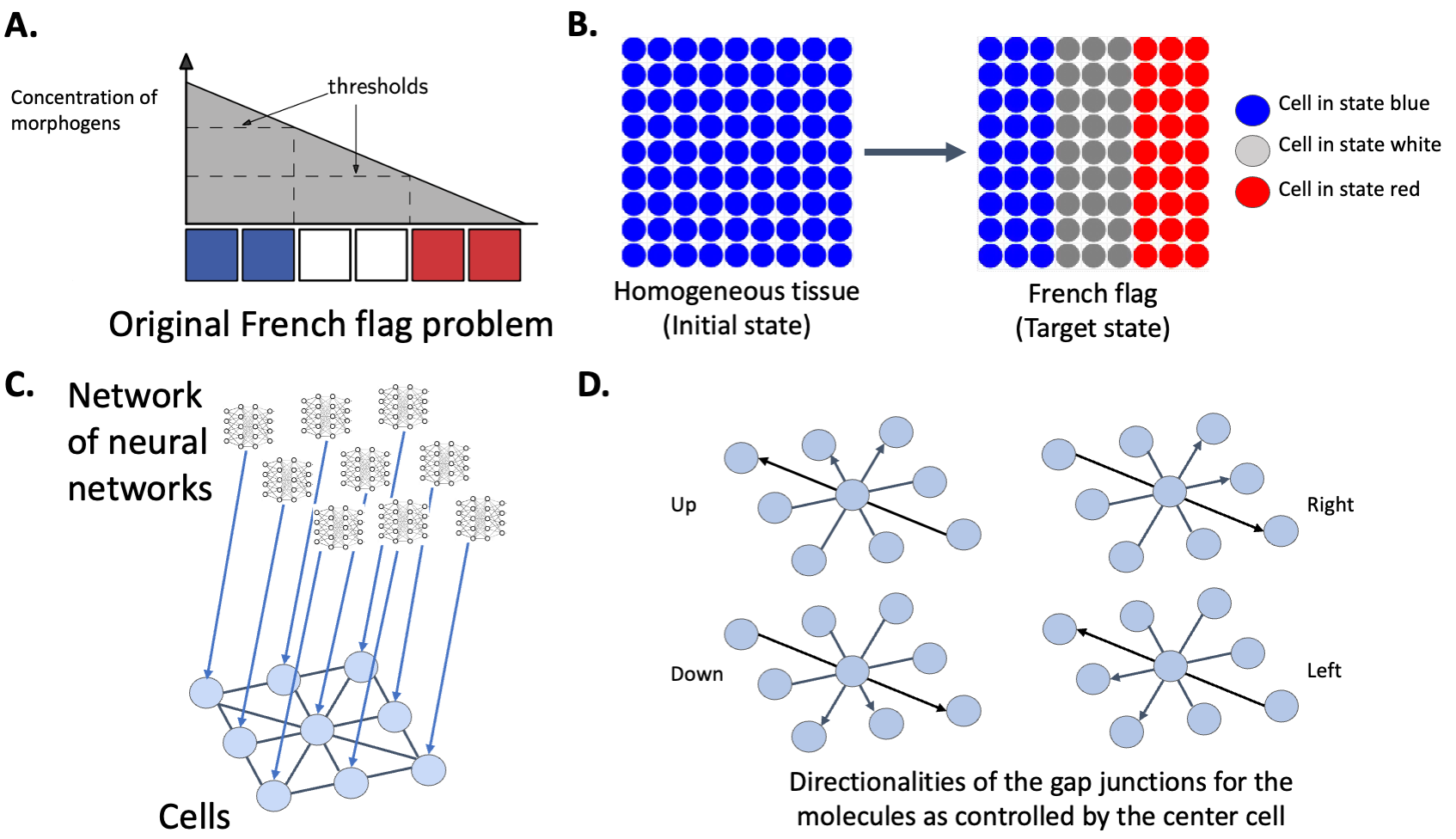}
  \includegraphics[width=\linewidth]{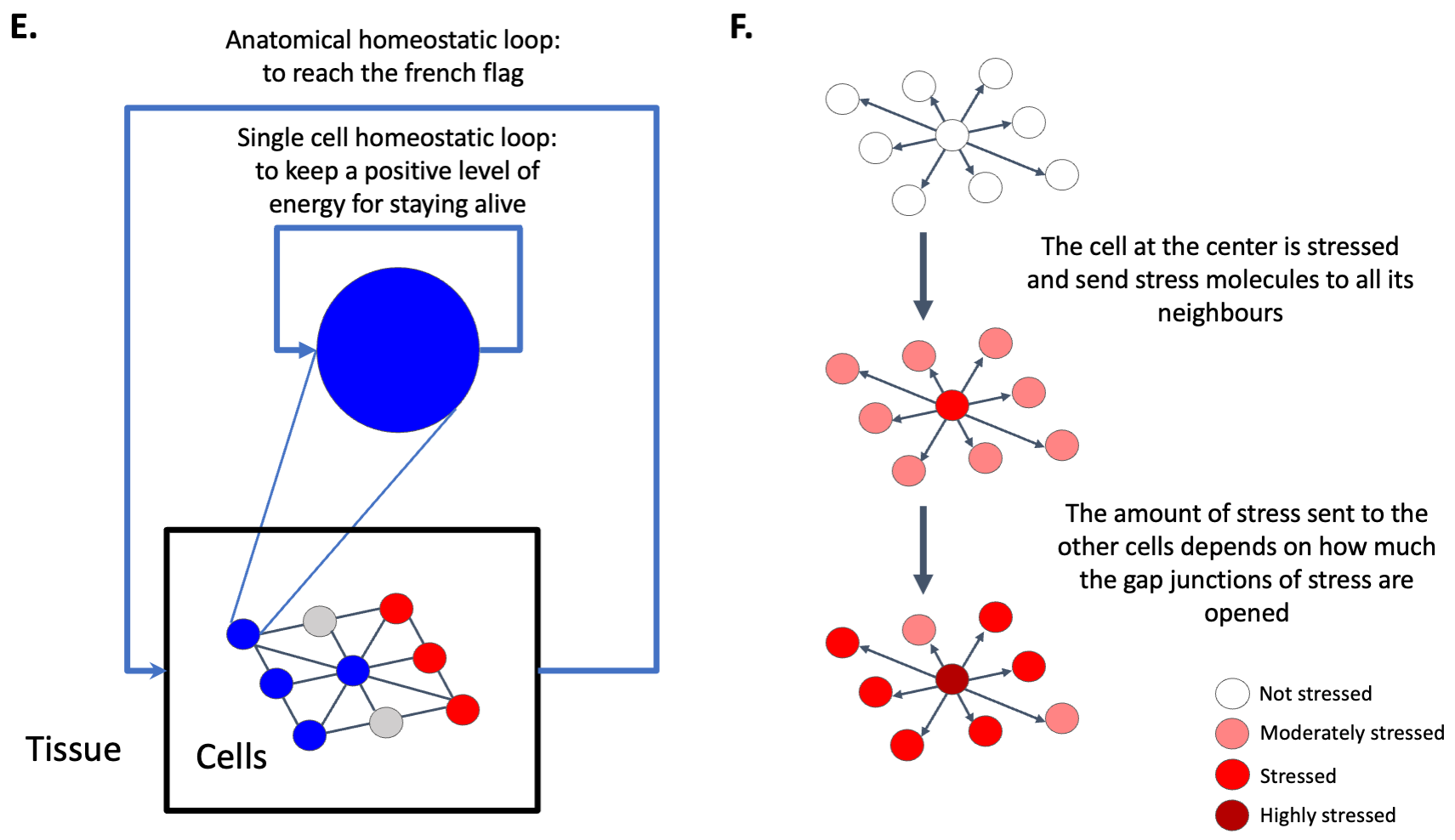}
  \caption{Description of the general scheme of the evolutionary system. \textbf{A.} Description of the French flag problem. The concentration of morphogen triggers a particular gene expression leading to 3 different states: blue, white or red. \textbf{B.} Description of the virtual environment. At left, the tissue is randomly initialized. At right, the tissue has reached the French flag. \textbf{C.} Each cell is connected to its neighbours and can exchange molecules via gap junctions. The behavior of each cell is controlled by its neural network. \textbf{D.} Directionalities of the gap junctions as controlled by the neural network embedded in the cells \textbf{E.} Two homeostatic loops have been implemented: a homeostatic single-cell loop, in which each cell has to maintain a positive level of energy to stay alive; and an anatomical homeostatic loop, in  which all cells receive energy depending on how closely the collective (or the tissue) approaches the anatomical goal. \textbf{F.} Spread of the stress molecule to neighbouring cells  as controlled by the ANN inside each cell. Alternatively, each cell can send to all its neighbours an anxiolytics molecule.}
  \label{fig:ff}
\end{figure}

We integrated a homeostatic loop that enables it to optimize for the levels of some particular chemical (we are not trying to model the origin of life itself – we start with cells that can keep themselves alive via basic homeostasis, e.g that need to keep a level of energy superior to 0).  We allow cells to attach to each other using gap junctions, to send intracellular signals and spread metabolites. These signals, and their properties – such as speed of spread, ability to propagate gap junctions, gating properties of gap junctions based on signal and cell state, etc. - are all coded by genomes through evolution. Cells don't know the origin of a given signal once it’s inside the cell; it has been suggested that this “wiping property”  enables a kind of shared memory that is essential for the scaling of small agents into larger intelligences \cite{levin2021technological}.  The cells have a minimal memory of the past and have access to their stress and energy levels, and their states, at time $t$ and $t-1$ (see Algorithm \ref{algo1}).

The general approach is related to cellular neural networks \cite{chua1988cellular} and more particularly to the growing neural cellular automata \cite{mordvintsev2020growing} where each cell has access to the states of its neighbours and contains a neural network to drive actions (neural networks are simply a convenient formalism for encoding cell behavioral properties, equal in power to gene-regulatory networks \cite{vohradsky2001neural}).  In our case, the number of cells is fixed and does not grow with time and the tissue formed by those cells has borders (i.e., it is not a torus).
\\\\
\noindent\textbf{Cell description.}
Each cell integrates an artificial neural network (representing the gene-regulatory networks and pathways operating within cells \cite{vohradsky2001neural, vohradsky2001neurala}) that controls the opening and closure of gap junctions in 4 directions: up, down, left, right. Artificial neural networks are commonly used to model genetic regulatory networks \cite{chai2014review}; in that line of research, the artificial neural network inside each cell can be interpreted as a genetic regulatory network controlling the metabolic processes. Each cell has one type of molecules that can go through the gap junctions depending on their openings and that will trigger different kind of genomic expression leading to three different cell states: blue, white, or red (see Figure \ref{fig:ff}1). These states depend on the level of particular molecules inside the cells (red if the level of molecules is between $0$ and $5$, grey between $5$ and $10$, blue if superior to $10$) and we imposed an energy cost on state change. Each cell also has a stress molecule and an antistress counterpart that can be sent to its neighbours (see Algorithm \ref{algo2}).

The use of the stress system is completely evolved as the number of molecules to be transferred is chosen by the embedded neural network. The neural network, which is identical for each cell, has $11$ inputs: the internal levels of molecules, energy at time $t$, energy at time $t-1$, stress at time $t$, stress at time $t-1$, the internal state at time $t$ and $t-1$, the size of the collective the cell is part of (number of cells of the same state connected by opened gap junctions in the tissue), the error between the size of the collective the cell is part of and how much it should size in percentage - e.g. the size of one stripe -  ($|100 - [\text{size of sub-collective}/(\text{total number of cells in the tissue}/3*100)]|$), the perception of the cell neighborhood (geometrical frustration or how similar the cell is to its neighbors) and a bias set at $0.5$. The 4 outputs of the ANN are: the number of molecules to send to neighbours, the number of stress molecules to send to neighbours and to be applied to the cell itself, the amount of stress-reduction molecules to send to neighbors and to be applied to the cell itself, and the opening of gap junctions in the 4 directions. The level of stress of one cell is bounded between $0$ and $100$.
\\\\
\noindent\textbf{Learning task.} The learning task is the French flag problem \cite{wolpert1969positional, wolpert2017French} (see Figure \ref{fig:ff}B). It was originally developed by Wolpert \cite{wolpert1969positional}, to formalize discussions about how spatial gradients might specify patterns of cell fates in a tissue. In this simulation, the spatial gradient that determines the cell state is the amount of molecules inside the cells. The two thresholds are 5 and 10 (if the number of molecules is superior to 10, the cell is in the blue state, between 5 and 10, it is grey, below 5, it is red).
\\\\
\noindent\textbf{From local to anatomical homeostasis.} We tied the single-cell and the anatomical homeostatic loops. Each cell has only one goal - to survive - and that corresponds to being in the appropriate state in order to receive energy (with the other members of the collective). On the other hand, the collective/tissue has a morphogenetic goal, which is to reach the French flag. 

\begin{algorithm}[t!]
\caption{Pseudo-code of agent.step()}
\begin{algorithmic}[1]
\Procedure{Agent.step(}{self, Input}       \Comment{Memory at t-1}
    \State self.energy(t-1) = self.energy
    \State self.state(t-1) = self.state
    \State self.stress(t-1) = self.stress
    \State output = self.communication()
    \State self.reward = Reward(self.position)
    \State self.energy += self.reward - 0.8
    \State UpdateStress()
    \If{self.energy <= 0}
        \State self.death()
    \EndIf
\EndProcedure
\end{algorithmic}
\label{algo1}
\end{algorithm}

\begin{algorithm}[t!]
\caption{Pseudo-code of agent.communication()}
\begin{algorithmic}[1]
\Procedure{Agent.communication(}{self, OutputANNMol, OutputANNStress}       

        \For{ i in directions}:
          \For{neighbour in neighbours}
            \If{self.GJ[directions[i]]*neighbour.GJ[i] $>$ 0}:   \Comment{If the corresponding GJs are opened}
                \State SendMoleculeNeighbour(i, OutputANNMol)
                \State neighbour.molecules += OutputANNMol*self.GJ[i]*neighbour.GJ[i]
            \State self.molecules -=OutputANNMol*self.GJ[i]*neighbour.GJ[i]
                \State neighbour.UpdateState()
            \EndIf
          \EndFor                             
          \EndFor                             
        
          \For{neighbour in neighbours}
            \If {self.GJStress $>$ 0}
                \State neighbour.stress+=OutputANNStress*((self.GJStress * neighbour.GJStress))   
                 \State  neighbour.stress-=OutputANNCalm*((self.GJStress * neighbour.GJStress))   
            \EndIf
          \EndFor 
          
          \If {self.GJStress $>$ 0}
             \State self.stress+=OutputANNStress
              \State self.stress-OutputANNAnxiolytics 
         \EndIf

        \State self.UpdateState()   \Comment{The level of molecules determines the state}

\EndProcedure
\end{algorithmic}
\label{algo2}
\end{algorithm}

At each step, each cell receives a reward in the form of an amount of energy  which is proportional to how closely its corresponding sub-collective (corresponding to one stripe and therefore bigger that the immediate neighborhood of one cell) is reaching the appropriate anatomical goal. Each cell therefore has a reward uncertainty, because the reward depends not only on on that cell's own behavior but also on the behaviour of the (sub)collective; in other words, the reward is affected by the decisions of distant cells. Therefore, the environment of the cells is of high uncertainty. In a sense, this scheme can be understood as a problem of multi-agent reinforcement learning under reward uncertainty. 

Goal-directed systems have the ability to focus on relevant information and ignore distracting information. To do so, they rely on  selective attention and/or interference suppression. Selective attention would rely on top-down biasing mechanisms as proposed by Desimone and Duncan \cite{desimone1995neural}. In our case, the top-down biasing mechanism is represented by the reward in energy that ties the two homeostatic levels. We also imposed an energy cost for communication and state changes (respectively of $0.8$ and $0.25$ per step).
\\\\
\noindent\textbf{Stress system.} Stress can be defined as three related concepts — the external and internal stimuli that cause stress, the emergency physiological and behavioral responses activated in response to those stimuli, and the pathological consequences of over-stimulation of the emergency responses \cite{le2007historical, romero2004physiological}. In our case, we added one more feature: a communication system that enables the diffusion of a stress molecule through the tissue to allow other cells to feel stress that was not caused by their own internal state. When the ANN makes the decision to send stress, each cell will diffuse a molecule to its neighbours that will increase equally the stress level of the stressed cell itself and its neighbours. In the same manner, the ANN can also cause the cell to send a stress-reduction  molecule (functionally equivalent to an anxiolytic intervention) to the neighbouring cells that will decrease the cell stress levels (see Figure \ref{fig:ff}E). Evolution can choose whether or not to use this communication system.
\\\\
\noindent\textbf{Evolutionary algorithm.} 
We use ES-Hyperneat \cite{risi2011enhancing} to simulate the evolutionary cycle. The fitness function is the percentage of good states the tissue has:

\begin{equation}
  fitness = (\text{number of good states}/\text{number of cells})*100
\end{equation}

The whole system has been coded in Python programming language, using the agent-based modeling framework Mesa \cite{masad2015mesa} and the MultiNEAT package. It is freely available on \url{https://github.com/LPioL/scalefreecognition}.  
\\\\
\noindent\textbf{Parameters.} 
All experiments in this article used the same parameters and have been repeated over $20$ runs. For each evolution, we used $250$ generations, a population size of $350$ individuals, a division and variance thresholds of $0.03$.  The energy cost for state change is $0.25$ and at each step, cells lose $0.8$ in energy. The minimum and maximum number of species are respectively $5$ and $15$. The number of generations without improvement (stagnation) allowed for a species is $10$. The depth for the neural net is $4$ hidden layers and $3$ for the quadtree.

The available CPPN activation functions for the French flag task domain were sigmoid, Gaussian, linear, sine, and step. The band-pruning threshold for all ES-HyperNEAT experiments was set to $0.3$. The bias value for the CPPN queries is $-1$.
The cells starts with an energy levels initialized at $70$. The energy cost of communication is set à $0.8$. We also applied an energy cost to the state change for all cells set at $-0.25$.


\subsubsection{Information-theoretic analysis}

Information theory \cite{mackay2003information} is a very useful tool to understand the information dynamics in complex systems. We used two information-theoretic measures to analyze the information dynamics of the results:  active information storage \cite{lizier2012local} and  transfer entropy \cite{schreiber2000measuring}.
\\\\
\noindent\textbf{Active information storage.} The amount of information in the past of one agent that is relevant to predict its future state is defined as the information storage. In this article, we focus on the active information storage (AIS) component, which is the stored information that is currently in use for computing the next state of the agent \cite{lizier2012local}. Formally, the AIS of an agent $Q$ is defined as the local (or unaveraged) mutual information between its semi-infinite past $q_{n}^{(k)}$ as $k\to \infty$ and its next state $q_{n+1}$ at time step $n + 1$:

\begin{equation}
    a_{Q}(n+1) = \lim_{k \to \infty} log_{2} \frac{p(q_{n}^{(k)}, q_{n+1})}{p(q_{n}^{(k)}) p(q_{n+1})}
\end{equation}

$a_{Q}(n, k)$ represents an approximation of history length $k$. The average
over time (or equivalently weighted by the distribution of $(q_{n}^{(k)}, q_{n+1}))$: $A_{Q}(k) = \left\langle a_{Q}(n,k) \right\rangle$. With AIS, an agent can store information regardless of whether it is causally connected with itself \cite{lizier2012local}.

In this article, we compute the local active information storage over the states of the cells.
\\

\noindent\textbf{Transfer entropy.} Transfer entropy is the information provided by the agent source about the destination’s next state that was not contained in the past of the destination agent. In this article, we use the local transfer entropy introduced by Lizier \cite{lizier2012local}. The local transfer entropy  from a source
agent $Z$ to a destination agent $Q$ is the local mutual information between the previous state of the source $z_n$ and the next state of the destination agent $q_{n+1}$,  conditioned on the semi-infinite past of the destination $q_{n}^{(k)}$ (as $k \to \infty$):

\begin{equation}
t_{Z \to Q}(n+1) = \lim_{k \to \infty } log_{2} \frac{p(q_{n+1} | q_{n}^{(k)}, z_{n})}{p(q_{n+1} | q_{n}^{(k)})}
\end{equation}

Transfer entropy $T_{Q}(n, k)$ is the (time or distribution) average: $T_{Q}(k) = \left\langle t_{Q}(n,k) \right\rangle$ and $t_{Q}(n, k)$ represents an approximation of history length $k$. While mutual information measures correlation only, the transfer entropy measures a directed and dynamic flow of information in the network of agents.

For the analysis of our information dynamics, we computed different kind of transfer entropy: the transfer entropy from stress (by discretizing it by chunks of $10$) to the states of the neighbours (blue, white, red). We computed the local transfer entropy pairwise with all the neighbours of one cell and averaged it.  In the same manner, we computed the transfer entropy from the energy state to the state of the cells and conversely the transfer entropy from the state the cells to the energy state.

We also defined the transfer entropy from one stripe to another as the sum of all transfer entropy of one cell of one stripe to all cells of the other stripes computed pairwise on the time series of the internal states. Formally, we define the averaged transfer entropy over a range of source-destination pairs in the spatial locations of different stripes $S_1$ and $S_2$:

\begin{equation}
t_{S_1 \to S_2} = \sum_{Z_i \in S_1}^{} \sum_{Q_j \in S_2}^{} t_{Z_i \to Q_i}
\end{equation}

The transfer entropy defined for specific subsets of tissue processes is
useful in considering distributed communications across agents with specific
roles.

\subsection{Method for the planaria experiment on repatterning}

\subsubsection{Colony maintenance}
Dugesia japonica were maintained in Poland Spring water at 20°C, fed calf liver paste once a week and cleaned twice a week, as described in \cite{oviedo2008establishing}. Animals were starved for one week prior to usage in amputation and pharmacological treatment experiments and were not fed for the duration of all experiments. 

\subsubsection{Animal manipulation}
Cutting of planaria was performed on a cooling plate using scalpel fragments. For generating headless animals, pre-tail fragments were cut by placing a cut at the pharynx opening and another cut narrowly above the tail tip. For tracking tail regeneration, animals were cut halfway between the head and tail. For tracking regeneration along the A/P axis, animals were either decapitated narrowly by cutting just below the auricles, cut halfway between the base of the head and the top of the pharynx, or cut directly at the top of the pharynx. To track regeneration of internal tissue a puncture wound was induced using square glass capillaries of a 0.7 mm2 inner diameter (VitroCom, Mountain Lakes, NJ) directly posterior to the pharynx opening. A fine paintbrush was used to remove the cut tissue from the middle of the animal. 

\subsubsection{Pharmacological Treatments}
Headless worms were produced through transient pharmacological treatment of freshly cut pre-tail fragments in 18 $\mu$M U0126 dissolved in DMSO (Sigma). No more than 40 fragments were treated per 10 cm petri dish. Fragments were incubated in U0126 for 3 days at 20°C before washing out the drug solution. At 14 days post amputation, the fragments were scored for regenerative phenotype. 

\subsubsection{Scoring of Repatterned Phenotypes}
Headless animals were maintained in individual wells of 12-well plates and their phenotypes were scored weekly for the duration of the experiment. Any significant changes in morphology, such as head regeneration, fissioning, or ectopic tissue development were noted. Time of repatterning was determines once one eye was visible in the forming head structure. Worms were labelled as “Polarity Flip” when a headless worm fissioned and a head regenerated at the posterior wound site. Worms were labelled as “Dorsal/Ventral repatterning” when an outgrowth of tissue occurred in the D/V plane of the headless worm. Worms were labelled as “Lateral growth” when significant changes in morphology occurred with growth on the lateral area of the animal, resulting in stable morphology which did not produce a head.

\section{Computational results}

We analyzed a number of experiments performed with this system, tracking key physiological parameters over time in each experiment (see Figures), including gap junctional communication, stress levels, and cell types as a function of position. We focused on the most biological evolved tissue: the one where stress is used as an instructive signal and increases and decreases in function of the homeostasis of the tissue.

\subsection{The tissue minimizes error for reaching the target morphology}

We first tested the ability of the cellular collective to solve the French flag problem, that is to organize the two-dimensional tissue into a 1-dimensional axis of positional information with respect to cell type identity. Each cell received energy according to is location on the tissue, and its energy reward was proportional to how well the other cells of one stripe of the French flag were resolving the French flag pattern (in other words, evolution gives partial credit for imperfect primary axial patterning, selecting for embryos with optimal morphogenesis). The cell colors in all figures represent cell fate, as in the original definition of the problem in which an embryo must spontaneously pattern itself into 3 ordered regions of cell fate \cite{wolpert1969positional}. All cells started in the blue state (homogeneous tissue corresponding to an un-patterned nascent blastoderm) and had to dynamically cooperate with, and send the appropriate signals to their neighbours, to solve the problem. The fitness function was computed for $100$ steps, defining the time course of development in this virtual embryo. 

\begin{figure}[!ht]
  \centering
  \includegraphics[width=0.7\linewidth]{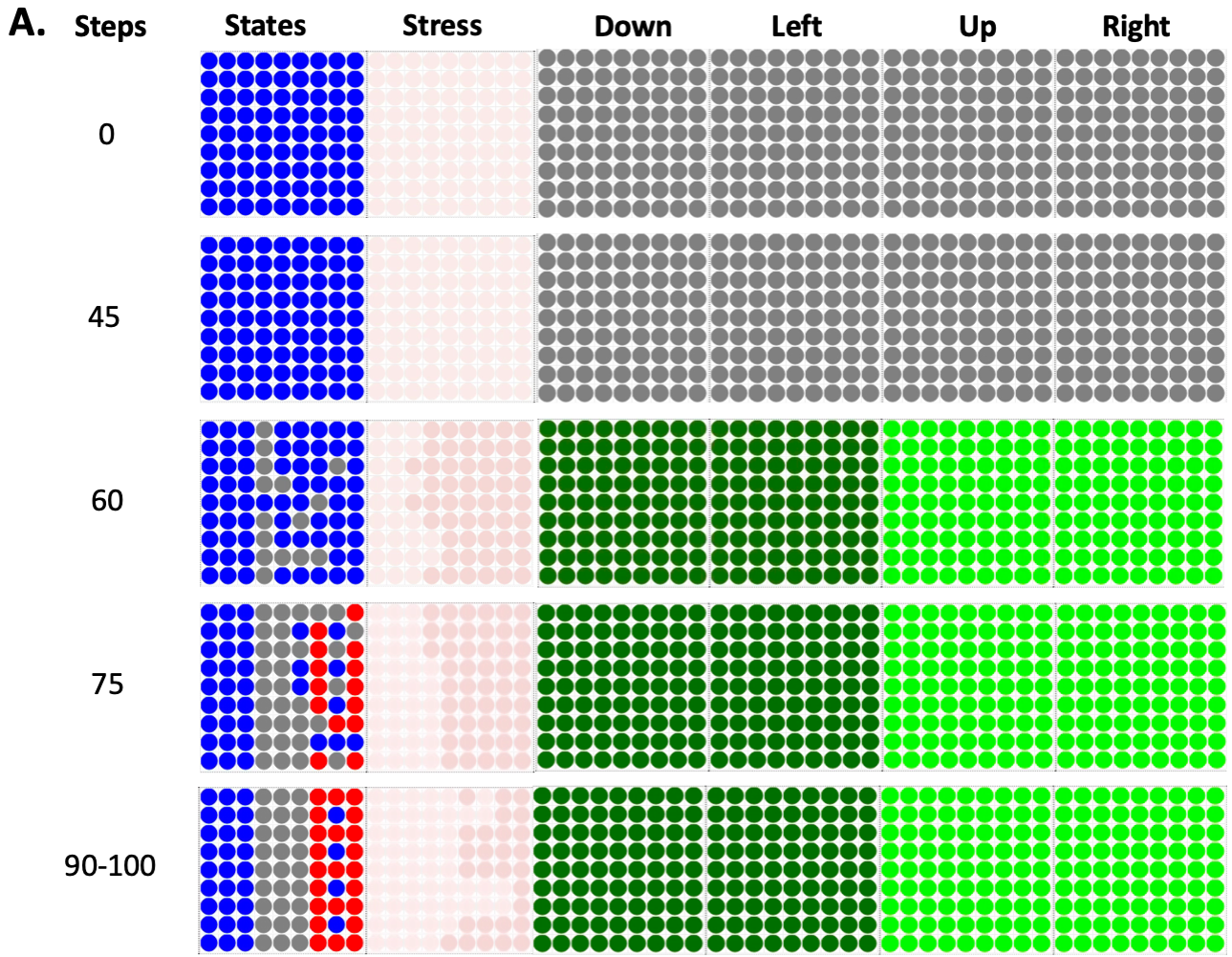}\\
  \includegraphics[width=0.7\linewidth]{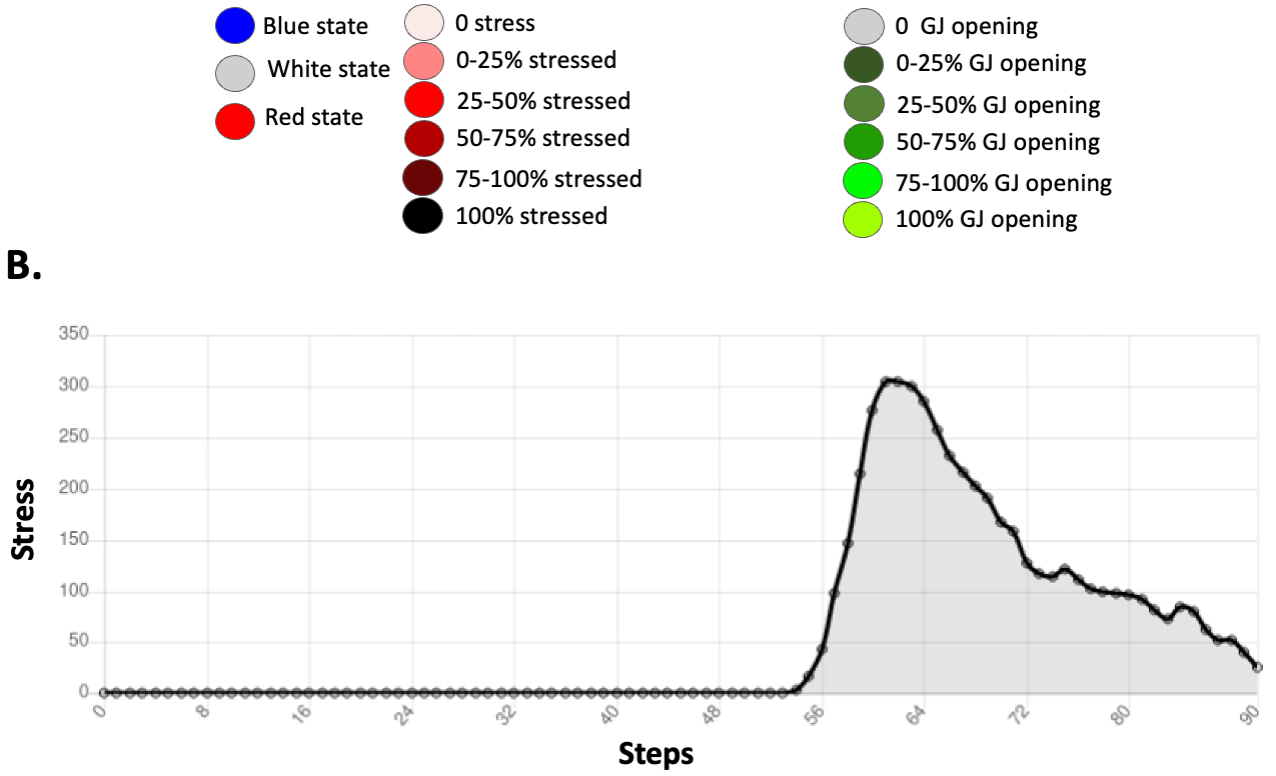}
  \caption{Dynamics of the resolution of the French flag problem. \textbf{A.} Time-lapses of the dynamics of resolution of the French flag problem by the tissue. The dynamics of the tissue is represented by 5  taken at 0, 30, 60, 90-100 steps of the dynamics of the tissue. The first column represent these steps, the second column represents the states of the cells in the tissue (blue, grey, red) and the  third column represents the use of the stress system on a red scale: the darker the red, the more the system is stressed.  The last four columns represent the opening in the gap junctions in $4$ directions: down, left, up and right. The darker the green, the less the gap junctions of each cells in the different directions are opened. The cells resolve the French flag problem almost entirely at the $90$ steps. The stress system is used all through the resolution of the French flag problem. \textbf{B.} Amount of stress inside the tissue during the resolution of the French flag problem. The stress increases around $55$ steps and decreases when the target morphology is reached.}
  \label{fig:ff_error_min}
\end{figure}

We observed that the ANN inside each cell evolved, and that this system was able to solve the problem (see Figure \ref{fig:ff_error_min}). A typical tissue behavior had the following features. First, the regions corresponding to future grey and red stripes began to be stressed and then it was mainly the red stripe that was stressed. All gap junctions were open, the upper and right gap junctions were fully open and the low and left gap junctions were half open. The gap junction states  weighted the flow of molecules so that  the stream from right  to left and right to left  were  equal  as was also the case for the streams in the vertical axis.  The diffusion of molecules  from left to right occurred  because the cells of the blue stripes  acted as a reservoir of molecules  and the cells decided   to send more or fewer molecules  to their neighbours  depending on the dynamics. 

In the representative example we show here, the stress  increased  at $55$ steps and it then decreased as the French flag morphogenetic  problem was resolved. At $90$ steps, the tissue reached $95.1\%$ of the French flag target morphology, and there were 4 remaining blue cells in the red stripe, (an almost perfect solution). The same result was seen over 20 repeat runs, solving the problem with axial quality rate of $94.4\% \pm 0.84$ over $20$ runs.

Thus, the tissue learned to minimize error between its current state and the target morphology during its lifetime in order to stay alive; by doing so, the tissue - starting from a homogeneous state - resolved the French flag problem. We conclude that this minimal system shows how cell behaviors (that can be tuned by evolution or learning) enable the collective to harness individual metabolic homeostatic loops (the pursuit of energy) toward a global patterning goal.

\subsection{The tissue is robust to perturbation}

The cells had learned  error-minimization in order to reach the anatomical goal, but did they follow a hardwired  plan that could only enable a feed-forward emergent pattern, or had they in fact acquired an ability for homeostasis  that would allow the tissue  to reach the target  morphology  in case of perturbations beyond the ab initio morphogenesis?

\begin{figure}[t!]
  \centering
  \includegraphics[width=0.7\linewidth]{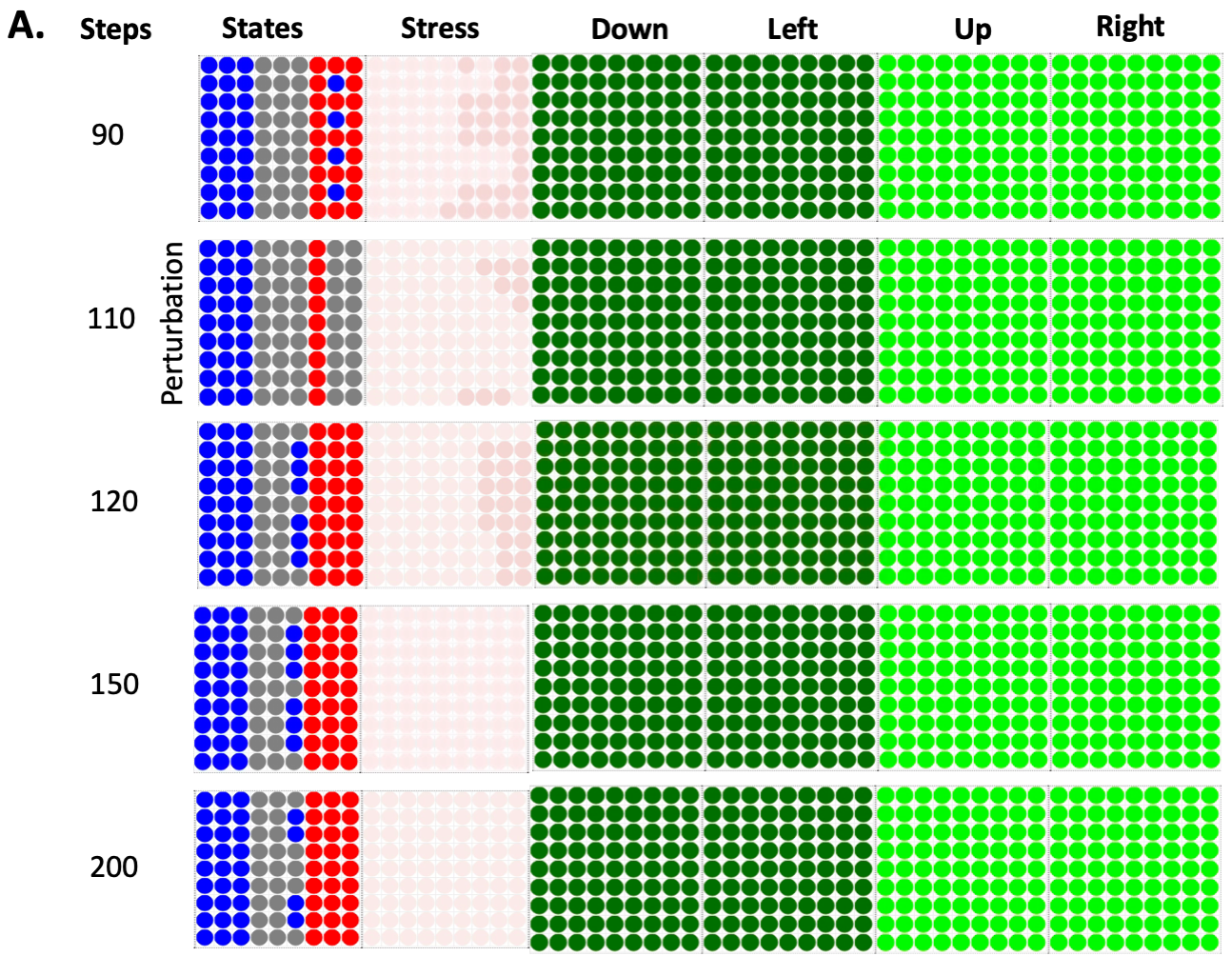}\\
  \includegraphics[width=0.7\linewidth]{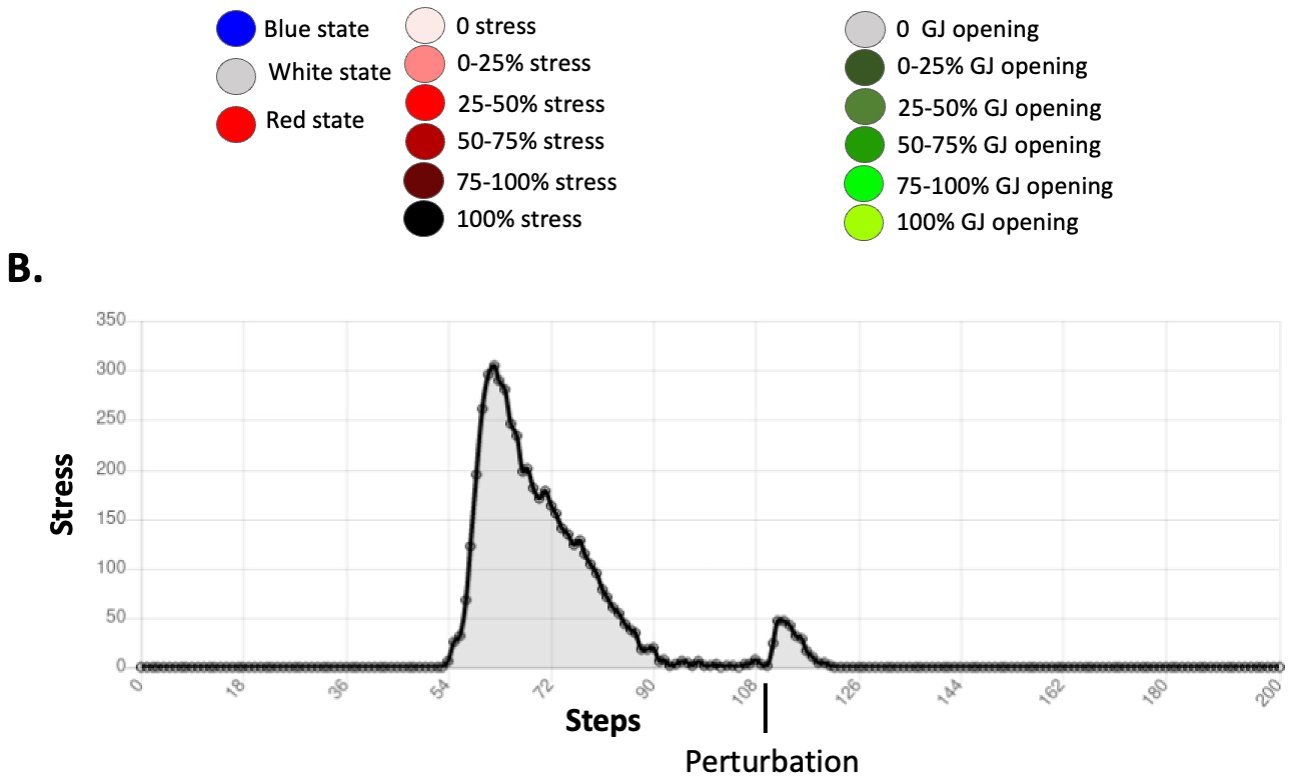}
  \caption{Dynamics of the resolution of the French flag problem after an induced perturbation. \textbf{A.} Time-lapses of the dynamics of resolution of the French flag problem by the tissue. The dynamics of the tissue is represented by 5  taken at 90, 110, 120, 150 and 200 steps of the dynamics of the tissue. The first column represent these steps, the second column represents the states of the cells in the tissue (blue, grey, red) and the  third column represents the use of the stress system on a red scale: the darker the red, the more the system is stressed. The last four columns represent the opening in the gap junctions in $4$ directions: down, left, up and right. The darker the green, the less the gap junctions are opened. The cells resolve the French flag problem almost entirely at $200$ steps after the artificial perturbation of the tissue at $110$ steps. \textbf{B.} Amount of stress inside the tissue during the resolution of the French flag problem. The stress system is used all through the resolution of the French flag problem and it increases and decreases with the resolution of the perturbation. }
  \label{fig:ff_homeo_loops}
\end{figure}

To determine the degree of plasticity of this process, in this next set of simulations, we perturbed the tissue by artificially changing at $110$ steps the states of the last two columns of the red stripe to grey cells. The cells had never been evolved on more than $100$ steps. However, after the perturbation at $110$ steps, the tissue corrected the red stripe in $10$ steps and a few blue cells appeared in the grey stripe (see Figure \ref{fig:ff_homeo_loops}). The stress increased and decreased in parallel to the perturbation and its resolution, as it did to resolve of the French flag problem. This capacity is similar to that observed in biological systems, many of which are able to regain normal morphology despite a wide range of perturbations \cite{pezzulo2016top}. The system tried to get rid of the blue cells left in the grey stripe. At $200$ steps, the tissue reached $95.1\%$ of the French flag. The dynamics of the gap junctions stayed the same as during the French flag resolution as seen in Figure \ref{fig:ff_error_min} without perturbation. 

Thus, we conclude that even though we did not specifically reward for anything other than one self-organizing property, in fact what the cells were able to do was repair to that setpoint from multiple starting configurations.

\subsection{The tissue maintains allostasis in adulthood}

\begin{figure}[t!]
  \centering
  \includegraphics[width=0.8\linewidth]{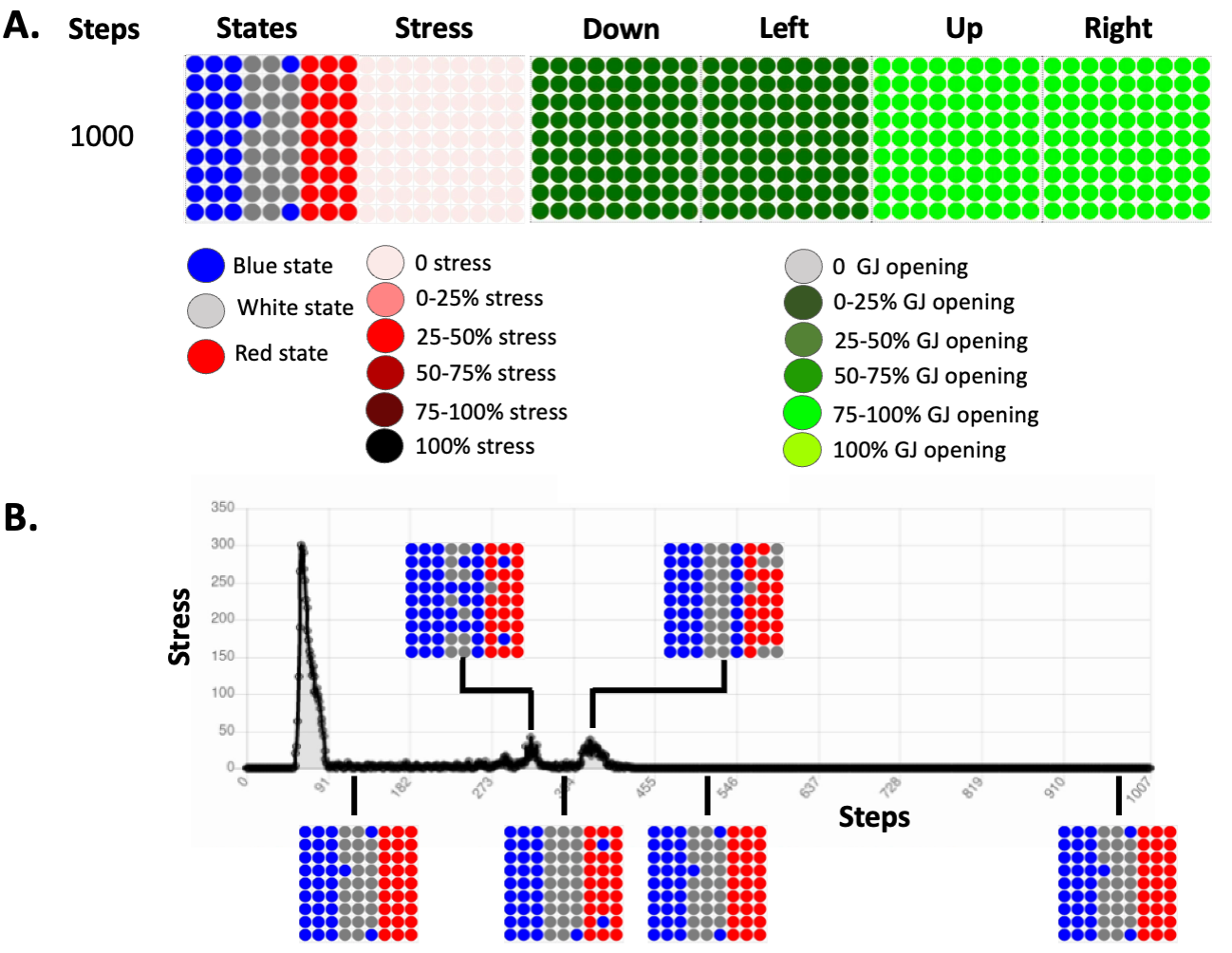}
  \caption{Long-term survival of the tissue and stress. The simulation was run for 1000 steps. \textbf{A.} State of the tissue at $1000$ steps. The first column represents these steps, the second column represents the states of the cells in the tissue (blue, grey, red) at $1000$ steps, and the  third column represents the use of the stress system on a red scale: the darker the red, the more the system is stressed. The last four columns represent the opening in the gap junctions in $4$ directions: down, left, up and right. The darker the green, the less the gap junctions are opened. At $1000$ steps, the tissue reached $96.3\%$ of the French flag target morphology \textbf{B.} Amount of stress inside the tissue during its lifetime. The stress increased and decreased when the tissue encountered deviations from the target morphology due to its intrinsic dynamics.}
  \label{fig:longterm}
\end{figure}

We discovered that the tissue can resolve the French flag problem and is robust to perturbation, but does the tissue reach long-term survival and maintenance of an adult phase ? To answer this question, we performed simulations running for $1000$ steps. We observed that the tissue maintained its morphology during the whole lifetime. In addition, at $1000$ steps, the morphology was even better than in the developmental phase, with the tissue reaching $96.3\%$ of the French flag target morphology on this simulation (see Figure \ref{fig:longterm} A.). The tissue was observed to spontaneously become stressed several times during its lifetime; these stress increases were not the consequence of any external perturbation, and were due to the intrinsic dynamics of the tissue. The first stress increase, as  described  above, happened prior to $90$ steps in order to reach enough of the target morphology to stay alive;  then we observed  (in the representative individual showed in Figure \ref{fig:longterm}B), $4$ different stress increases at $285$, $322$, $395$, and $896$ steps. At these steps, the intrinsic dynamics of the tissue created deviations from the target morphology that increased the level of stress, after which  the tissue  corrected the new anatomical trajectory in order to reach homeostasis, at which point  stress decreased. Once the tissue reached a morphology compatible with life, stress was reduced to $0$. On average for this task, the collective reached $87.7\% \pm 13$ of the target morphology over $20$ runs.

Interestingly, the cells had never been evolved over a period of time greater than $100$ steps – there was no selection pressure for long-term stability or survival. However, they were able to maintain the tissue more than $10$ times longer than their initial developmental period (the only aspect on which they were evolved). We were also surprised to find they also learned allostasis (following  the definition of McEwen and Wingfield \cite{mcewen2003concept}: allostasis is the process of maintaining stability (homeostasis) through change in both environmental stimuli and physiological mechanisms). In this simulation, the tissue changed regularly as it tried to get rid of remaining, inappropriately-located blue cells;  sometimes this remodeling process took the collective a sufficient distance from  the French flag target morphology to activate homeostatic mechanisms which then drove it back to normal, allowing long-term maintenance and survival of the tissue. 

Thus, we conclude that long-term stable survival does not need to be specifically selected for, as intrinsic dynamics and emergent anatomical regenerative capacity are enough to maintain order through an adult phase. Homeostasis is an active process which is mirrored in this case by the level of stress that increases and decreases with the distance between the tissue and the target morphology.

\subsection{Stress: different use-cases}

The morphogenetic process exhibits interesting stress dynamics, consistent with the proposals \cite{levin2021technological} that homeostatic loops are driven by stress as a reflection of delta from setpoint, and  that complex setpoints such as tissue-level morphogenetic patterns could arise from cells sharing stress information to optimize plasticity and coordinate in more complex problem spaces. Thus, we next studied the functional role of stress in the emergent morphogenesis we observed.

\subsubsection{Evolution exploits stress as an instructive signal to reach the target morphology}

\begin{figure}[ht!]
  \centering
  \includegraphics[width=0.7\linewidth]{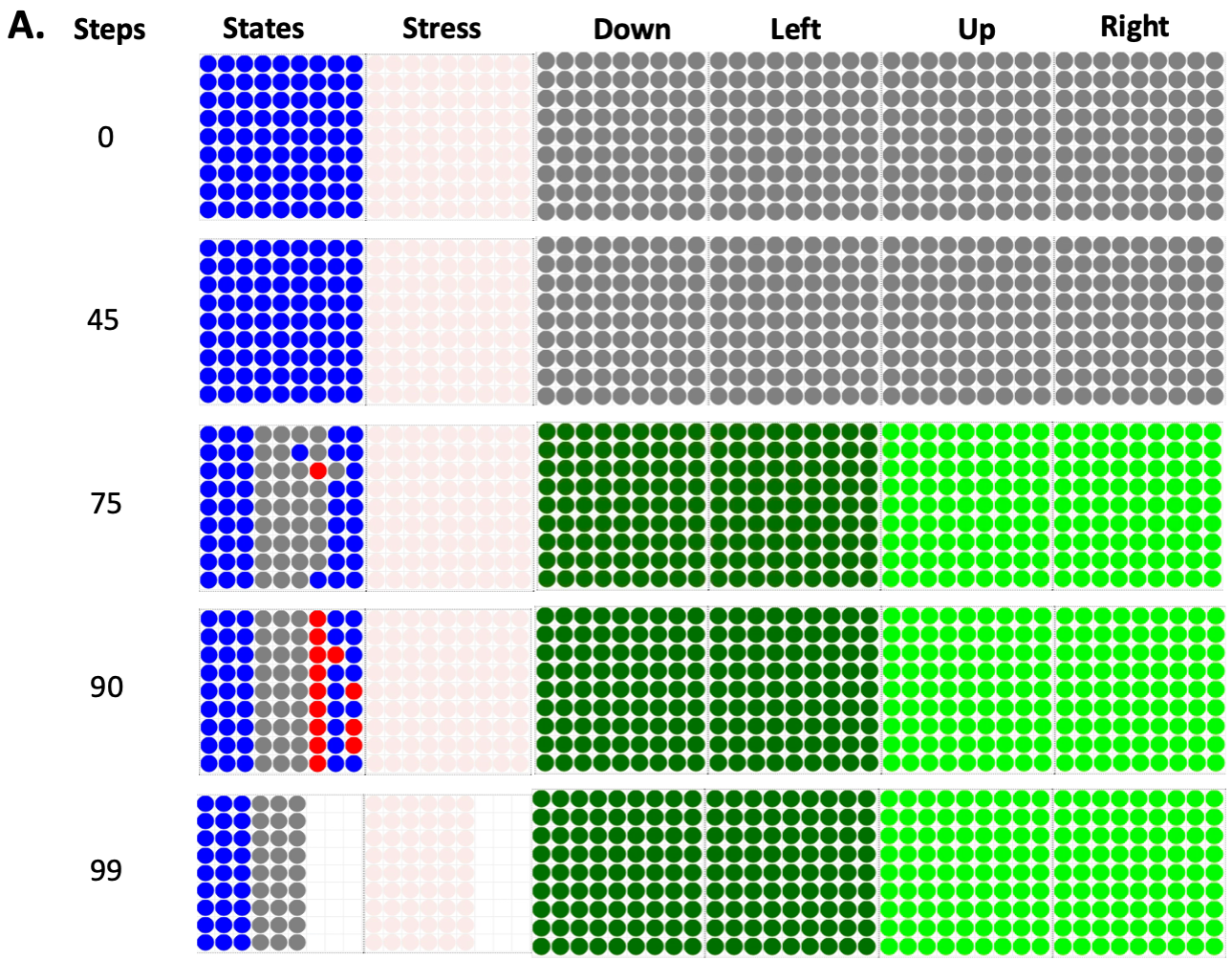}
  \includegraphics[width=0.7\linewidth]{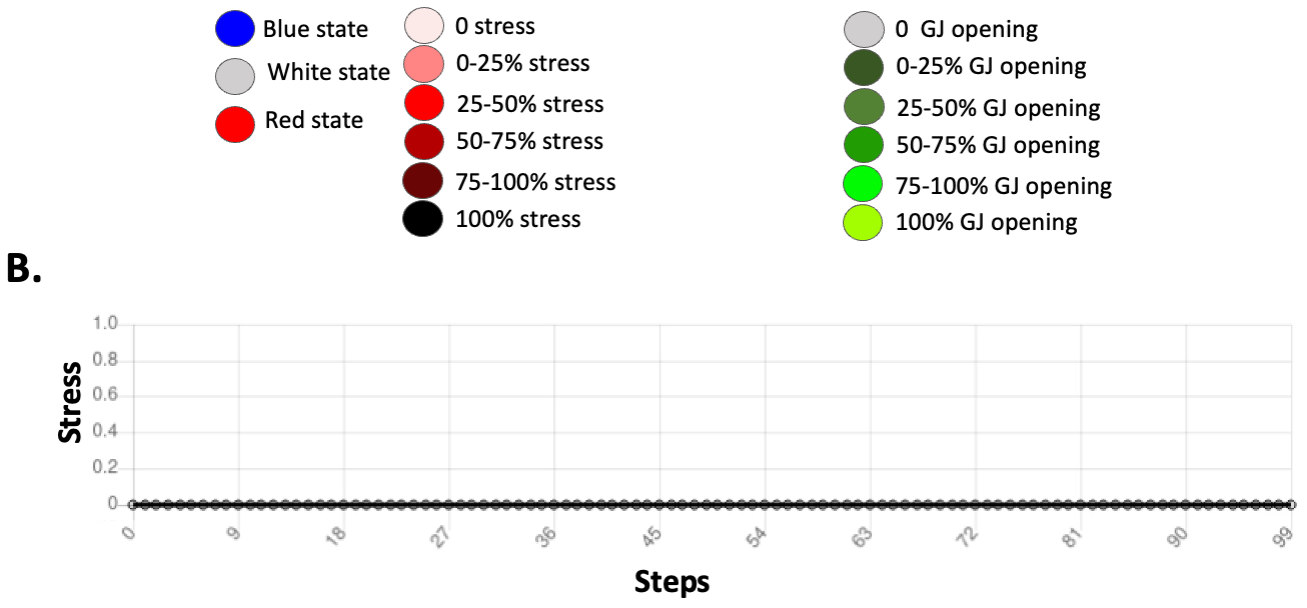}
  \caption{Dynamics of the resolution of the French flag problem with artifical injection of anxiolytics in the tissue. \textbf{A.} Time-lapses of the dynamics of resolution of the French flag problem by the tissue. The dynamics of the tissue is represented by 5  taken at 0, 45, 75, 90 and 99 steps of the dynamics of the tissue. The first column represents these steps, the second column represents the states of the cells in the tissue (blue, grey, red), the  third column represents the use of the stress system on a red scale: the darker the red, the more the system is stressed. The last four columns represent the opening in the gap junctions in $4$ directions: down, left, up and right. The darker the green, the less the gap junctions are opened. The cells behave completely differently without stress, at $90$ steps, the tissue had not reached the French flag configuration and at $99$ steps, the cells located on the red stripe died. \textbf{B.} is artificially held steady  at $0$ with simulated anxiolytics (a loss-of-function experiment for the stress system).}
  \label{fig:ff_anxiolytics}
\end{figure}

We first sought to determine whether stress  was  instructive or merely a byproduct of the various dynamics. To answer  this question,  we simulated a loss-of-function of the stress system in the tissue (e.g., as the action of an anxiolytic drug),  which forcibly reduced  the level of stress  to $0$ during the whole lifetime of the embryo  (see Figure\ref{fig:ff_anxiolytics}).

The dynamics  of the gap junctions  were similar in this anxiolytic experiment as they were  when stress  was used  (see Figure \ref{fig:ff_error_min}). However, we observed that at $90$ steps, the tissue reached only $82.8\%$ of the target morphology. And at $99$ steps, the collective of cells corresponding to the location of the red stripe of the French flag dies. The collective achieved $67.3\% \pm 2.8$ of the target morphology over $20$ runs.

These experiments reveal that anatomical homeostasis is not reached nor maintained during the lifetime of the cells without stress. In contrast, when the stress is available to the cells, at $90$ steps, the tissue reaches $95,1\%$ of the French flag target morphology and it maintains the morphology after this step.

Thus, we conclude that the ability to use the stress mechanism is functionally exploited by cells to implement morphogenesis and survival of development.

\subsubsection{Stress is instructive but only at particular levels}

\begin{figure}[!ht]
  \centering
  \includegraphics[width=0.7\linewidth]{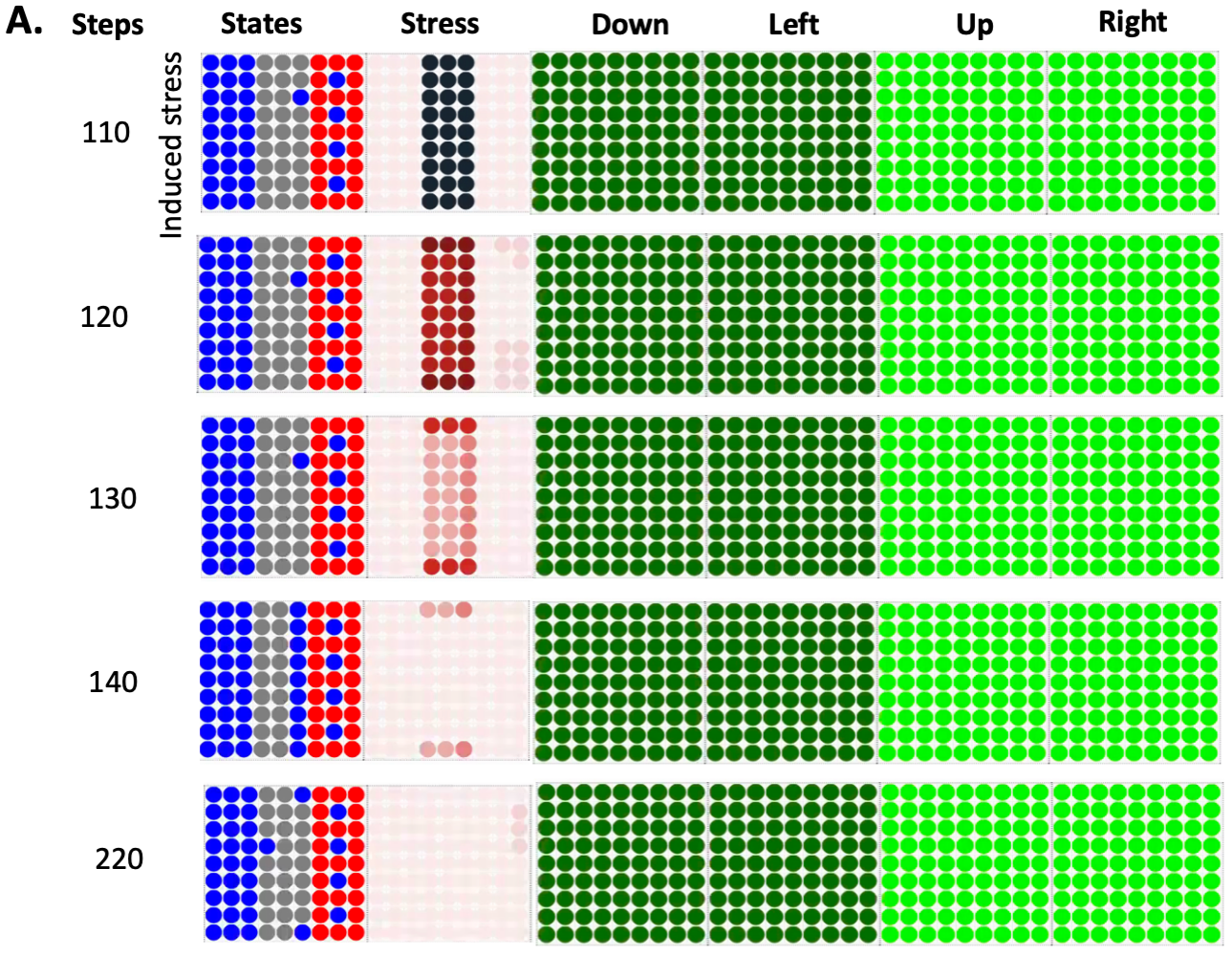}
  \includegraphics[width=0.7\linewidth]{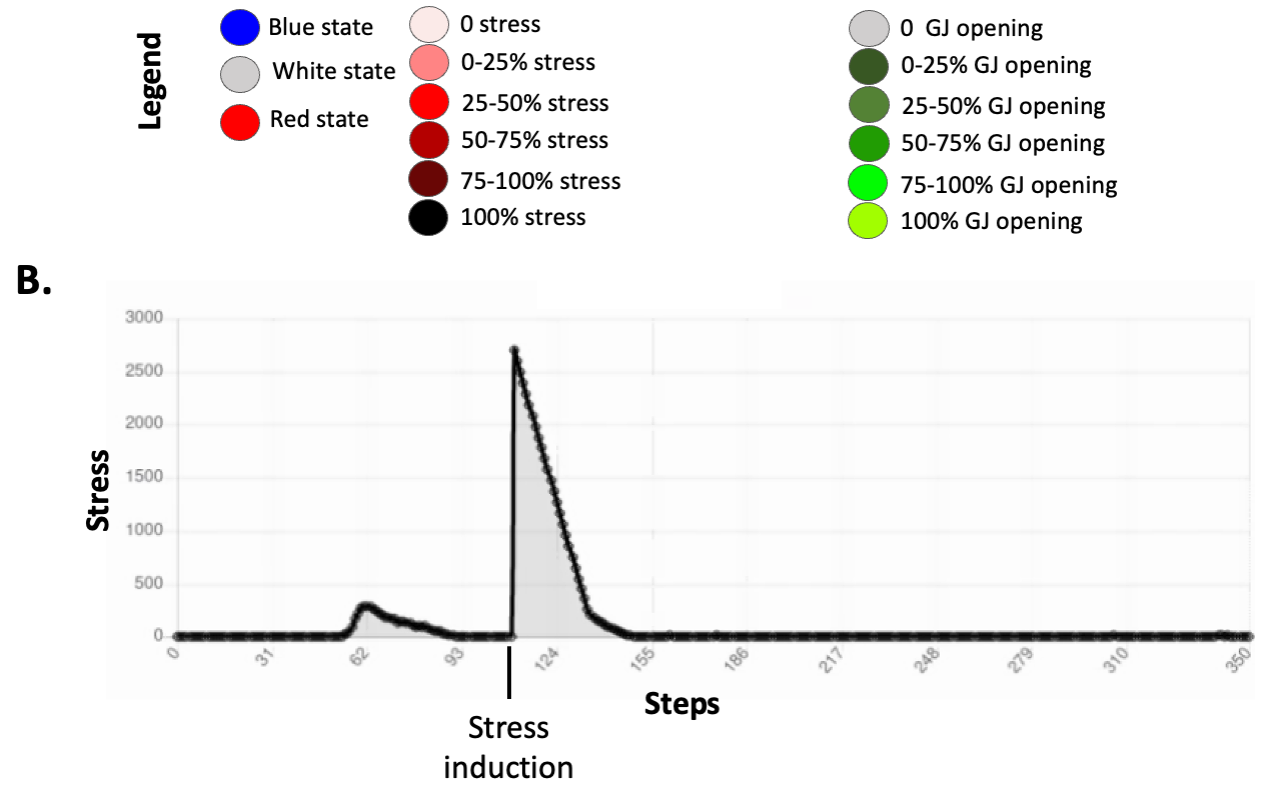}
  \caption{Dynamics of the resolution of the French flag problem after an artificial increase of stress. \textbf{A.} Time-lapses of the dynamics of resolution of the French flag problem by the tissue. The dynamics of the tissue is represented by 5 timesteps taken at 110, 120, 130, 140 and 220 steps of the dynamics of the tissue. The first column represent these steps, the second column the states of the cells in the tissue (blue, grey, red). The  second column represent the use of the stress system on a red scale.  The last four columns represent the opening in the gap junctions in $4$ directions: low, left, up and right. The darker the green, the less the gap junctions of each cells in the different directions are opened. At $140$ steps, it appears a wall of blue cells in the grey stripe, it seems the stress is informative but only at particular concentrations \textbf{B.} Amount of stress inside the tissue during its lifetime. The stress is artificially increased drastically at $110$ steps. Then, it decreases linearly until it reaches $0$.}
  \label{fig:induced_stress}
\end{figure}

Having seen the importance of stress mechanisms among cells, we next wanted to know if the stress system featured a window of optimality: is too much stress bad for this process, as it is for real biological systems?   Thus, we artificially induced a very high stress (compared to what is needed and used during the French flag resolution) at the locations of the grey stripe at $110$ steps of the simulation (see Figure \ref{fig:induced_stress}). With this excess stress, we simulated an environmental stressor that is not adaptively informative about the current situation of the tissue as it had already reached a version of the French flag compatible with survival (e.g. the cells receive enough energy). Stress decreased immediately without to change the states of the cells. At steps 140, a line of blue cells appeared in the grey stripe. It seems that stress was only instructive at  specific concentrations inside the cells, and then, blue cells appeared in the grey stripe. Stress transformed the cells close to the borders of the stripes into reservoirs of molecules, and by accumulating molecules, they became blue. It seems the cells evolved by making stress instructive only at specific levels as it had no effect during those last $30$ steps and it kept decreasing.

We conclude that evolution used stress as a communication system using a particular encoding, here the level of stress is instructive and have effects on the cell behavior only at specific concentrations.

\subsubsection{Evolution exploits the stress system but anxiolytics do not always impact  anatomical homeostasis}


Evolution finds more than one way to solve a problem. Here, we analyzed the counterexamples – the individuals that can solve the problem without the use of stress. We found that in $10$ evolutions that achieved a fitness score superior to $90\%$ on average over $20$ runs for reaching the French flag, we obtained $93\% \pm 2.6$ of reaching of the target morphology and $91.5\% \pm 8.$ when anxiolytics are added. However, $70\%$ of the evolved tissues use the stress system but it is not systematically functional. This high standard deviation for the anxiolytics experiment is explained by the fact that for the larger part of the evolved individuals, stress is not fundamental to reach the target morphology, for one evolved tissue its absence results in a dying embryo during development, while for others it has no negative impact. Quantitatively, 1/10 of the evolved tissues used stress as an instructive signal and die under anxiolytics (e.g. the one we presented the results on the different experiments above). This ratio is low but in terms of biological evolution, it would be enough to spread in entire the population over time if it increases the overall fitness of the organism during evolution. The other virtual embryos can develop the French flag without the need of stress even if they  used it as a result of  artificial evolution. 3/10 of the virtual embryos had better score on average on the French flag with anxiolytics, meaning that stress is a noisy signal for them.

\begin{figure}[t!]
  \centering
  \includegraphics[width=0.5\linewidth]{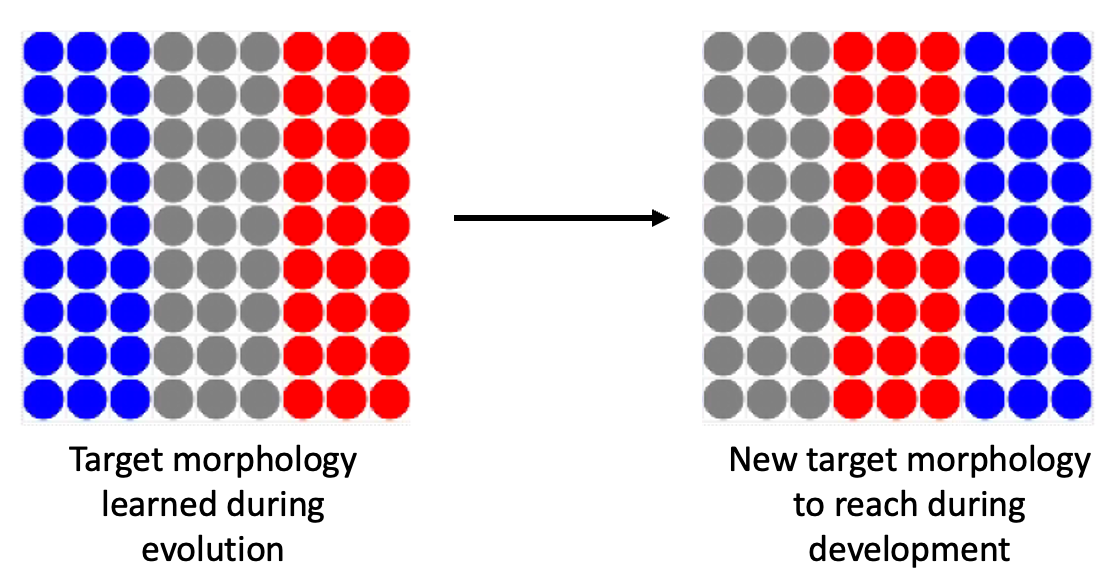}
  \caption{Representation of the target morphology used learned during evolution e.g. the French flag (left) and the new target morphology the tissue has to reach at $60$ steps (right).}
  \label{fig:new_target} 
\end{figure}

Why does biological evolution use stress -  does the use of the stress system in our simulation increases the adaptive responses of the organisms in more complex scenarios ? In order to answer this question, we tested the 10 virtual evolved tissue  achieving a fitness score superior to $90\%$ on average over $20$ runs for reaching the French flag on an experiment where the target morphology changes during the development phase (at $60$ steps, see Figure \ref{fig:new_target}) and cells that die are immediately regenerated by grey cells. We obtained an mean score of $74\% \pm 17.4$ on the new target morphology for the $10$ virtual embryos. For the one using the stress system as functional, we obtained a mean score of $89.2\% \pm 4.5$ and for the 9 others that don't use stress as instructive the mean is $73\% \pm 18.6$. We applied a one sample t-test to compare these two means and found the difference statistically significant (with $\text{p-value} = 0.0312$). It may be possible that stress enables adaptive responses in potentially harmful  environments. 

In summary, we found that the stress system is not always needed to resolve the French flag problem during development. Anatomical homeostasis without stress is possible on the French flag problem but exploiting spatial propagation of stress information  seems to increase adaptive fitness by improving performance in  scenarios where the normal process of morphogenesis can be impacted by external perturbations. As has been suggested \cite{levin2021technological}, stress is an ideal parameter for evolution to exploit as a representation of the error in homeostatic contexts, and the drive to reduce stress could then implement “grow and remodel until complete” in regenerative settings \cite{levin2021technological}.



\subsection{Information-theoretic analysis of the dynamics of the anatomical homeostasis}

\begin{figure}[ht!]
  \centering
  \includegraphics[width=\textwidth]{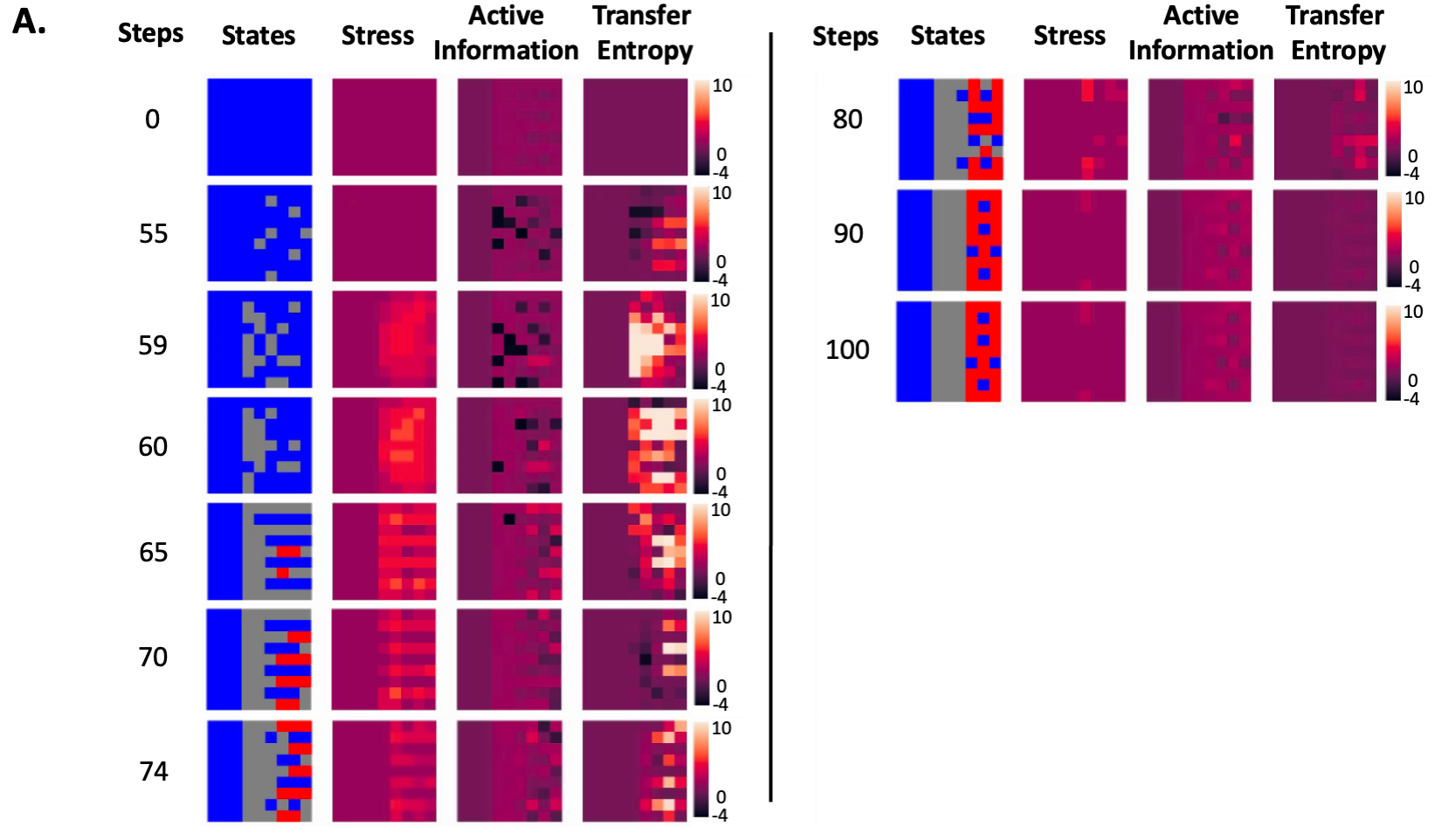}\\
  \includegraphics[width=\textwidth]{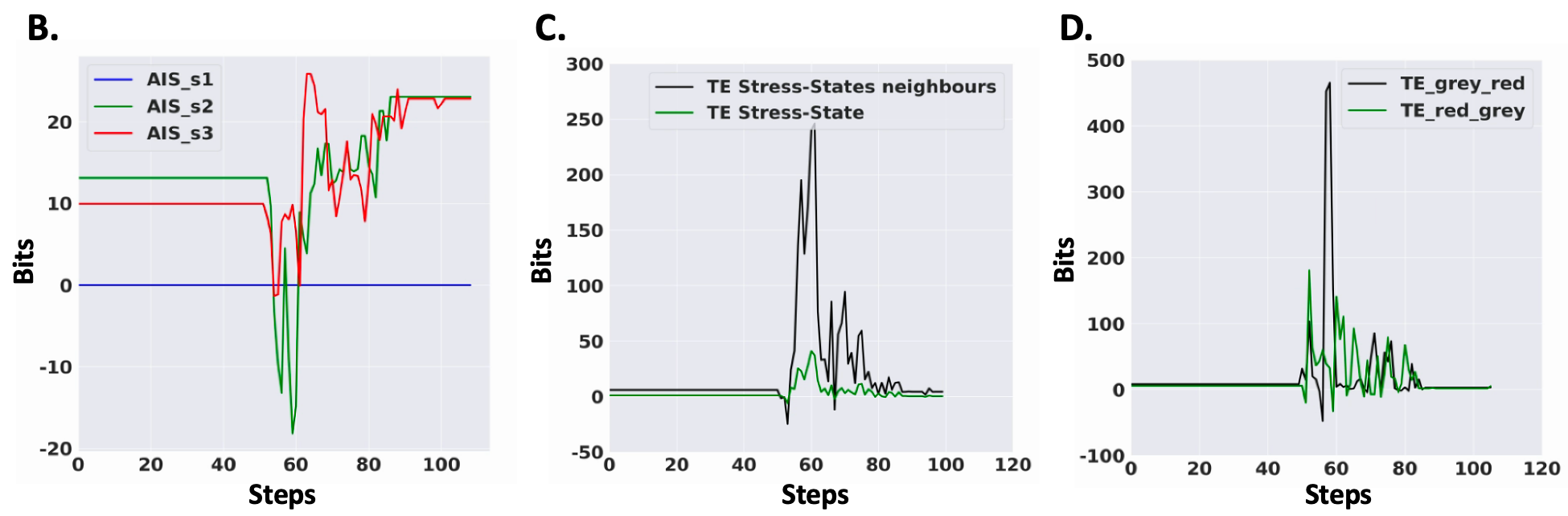}
  \caption{Information-theoretic analysis of the dynamics of the tissue during the resolution of the French flag problem. \textbf{A.} Time-lapses of the dynamics of resolution of the French flag problem by the tissue. The dynamics of the tissue is represented by 10 time-lapses taken at 0, 55, 59, 60, 65, 70, 74, 80, 90 and 100 steps. The first column represents these steps, the second column represents the states of the cells in the tissue (blue, grey, red), and the third represents the use of the stress system on a red scale. The last two columns represent active information and transfer entropy from stress to state of the neighbours. \textbf{B.} This figure represents the sum of local active information storage (AIS) by stripes, s1 is the blue stripe (blue), s2 is the white stripe (green), and s3 is the red stripe (red). AIS peaked first for the red stripe. \textbf{C.} This figure represents the sum of the  averaged local transfer entropy from stress time series to the state time series of the neighbors (black) and the transfer entropy from stress states to the internal states of the cells (green). The transfer entropy was much higher for the first one. \textbf{D.} This figure represents the transfer entropy from one stripe to another: grey to red stripe (black), and red to grey (red). The transfer entropy from one stripe to another peaked at different times suggesting that they alternated in driving the changes.}
  \label{fig:ff_info_analysis}
\end{figure}

To truly understand how cellular collective behavior solves morphogenetic problems, and to derive efficient interventions for biomedical contexts, it is essential to understand what the cells are saying to each other: what information is available to groups of cells, and how its propagation through the tissue enable the cohesion of multiple competent agents into an emergent individual at a higher level of organization that achieves a morphogenetic goal. Thus, we next applied information-theoretic measures to analyze the dynamics of the tissue during anatomical homeostasis. We computed the local active information and local transfer entropy for the original simulation without any perturbation (see Figure \ref{fig:ff_error_min}). We also computed the transfer entropy from one stripe to another to understand if one sub-collective could drive the information dynamics and ultimately the anatomical homeostasis (see Figure \ref{fig:ff_info_analysis}). 

\begin{figure}[!ht]
\centering
  \includegraphics[width=0.8\linewidth]{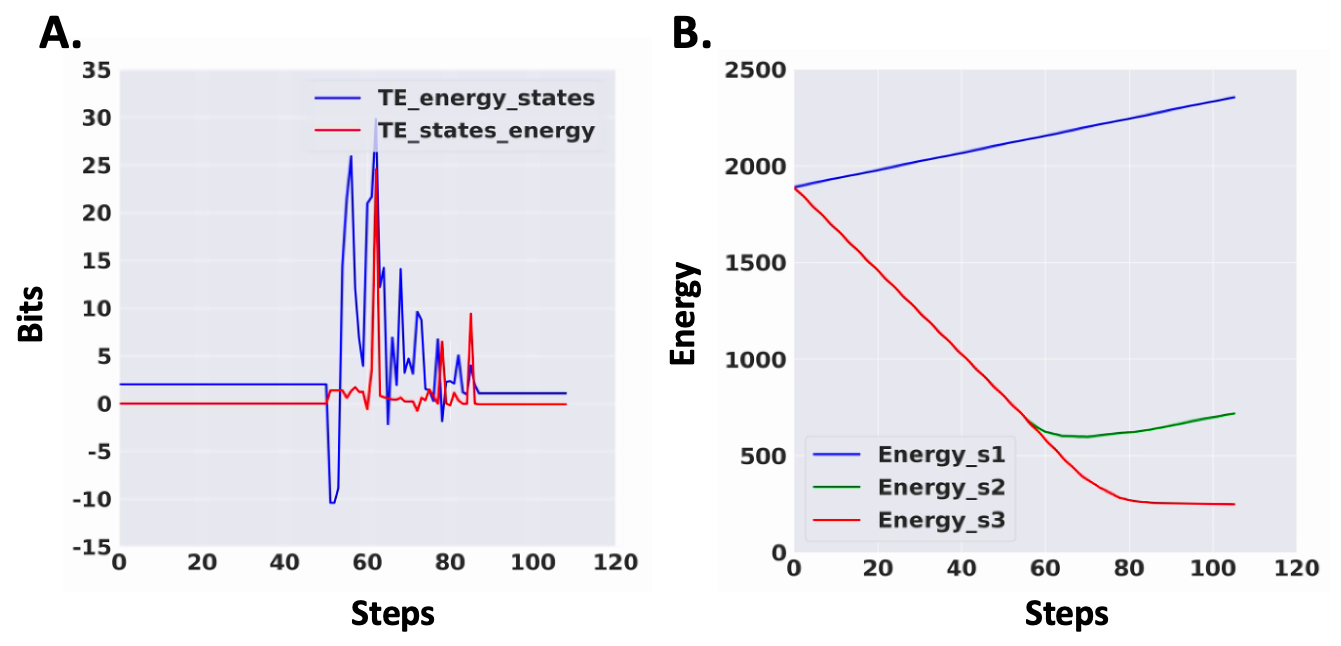}
  \caption{Information-theoretic analysis of the multi-scale homeostasis during the resolution of the French flag problem. \textbf{A.} This figure represents the transfer entropy from energy to states (blue) and the transfer entropy from state to energy (red). The first one peaks first suggesting that the information flow was coming primarily from the anatomical homeostatic loop. \textbf{B.} This figure represents the amount of molecules by stripes. The blue stripe is represented in blue, the grey stripe in green, and the red stripe in red. When the French flag configuration was reached, the cells received energy.}
  \label{fig:energy_states}
\end{figure}

The transfer entropy from stress of one agent to its neighbour started to increase in the tissue at timestep $55$. Then we wave an important increase at timestep $59$ in the locations of the grey and red stripes. This correlates with the increase of stress in the same spatial locations. The overall transfer entropy was almost $300$ bits at this timestep (see Figure \ref{fig:ff_info_analysis}C). Then, the transfer entropy decreased and reached a steady-state around $0$. The black squares represent negative local transfer entropy, meaning that the cells receive information from the stress that is misleading for predicting their next state. We also compared the transfer entropy from stress of one cell to its own state and found it to be much less important than the transfer entropy from the stress of one cell to the states of its neighbours (see Figure \ref{fig:ff_info_analysis}C). This suggests the key signaling mode for the stress is paracrine, not autocrine.

The dynamics of the local active information storage on the cell states of the cells followed a different dynamics. When transfer entropy was increasing, we could observe several spatially located negative local active information storage among the  spatial locations of the grey and red stripes (see Figure \ref{fig:ff_info_analysis}A). At $55$, $59$ and $60$ timesteps, the AIS was mainly negative, after which it would increase gradually for the grey until it reached the steady-state (see Figure \ref{fig:ff_info_analysis}A and \ref{fig:ff_info_analysis}B). For the red stripe, AIS was also globally decreasing and then we observed an important increase around $70$ steps, the cells could predict better their future state from the past memory. Concurrently, the stress is decreasing. The local AIS of the blue stripe was $0$ all along the dynamics as the cell never changed their states. Thus, stress is increasing concurrently with a decrease in local active information storage in the tissue as seen in Figure \ref{fig:ff_info_analysis}B) or in other words when the memory in the past is not effective to predict the future. This is compatible with the definition of stressors as unpredictable and/or uncontrollable stimuli \cite{levine1991id}. 

We also computed the average local transfer entropy from one stripe to another,  from the cells corresponding to the spatial locations of the red stripe to the grey and conversely (see Figure \ref{fig:ff_info_analysis}D). This averaged transfer entropy was $0$ at each time step for the blue to red and grey stripes and oppositely. With a time window of $k=4$, before $60$ steps, the tissue change seemed driven by the red stripe as we have a peak of $250$ bits of transfer entropy from the red stripe to the grey. Just after $60$ steps, it became the opposite, transfer entropy was much higher from the grey stripe to the red with a peak around $450$ bits of information suggesting that it was now the grey stripe that was driving the change in the states. The 'organizer' region seemed to change over time: embryonic regions take turns driving the large-scale process.

Finally, we computed the average of the local transfer entropy from energy to the state of one cell for the whole tissue and conversely from the cell state to energy (see Figure \ref{fig:energy_states}A). The first peak  we observed around $55$ steps corresponded to the transfer entropy from energy to states. The information flow came from the anatomical homeostatic loop then the two transfer entropies overlapped after $60$ steps, and at the same moment the averaged energy of all stripes was increasing or stopped decreasing (see Figure \ref{fig:energy_states}B). In our scheme, the information was first driven by the anatomical loop, and the energy that was needed for single-cell survival was controlled by this loop.

We conclude that stress was very informative in the state changes of direct neighbours of one cell. Stress was increasing concurrently with a decrease in local active information storage in the tissue as seen in Figure \ref{fig:ff_info_analysis}B) similarly to biological system where stressors are defined as an unpredictable stimuli. In our scheme, evolution used the stress system as a communication system that was activated when the memory in the past was not effective for predicting the future. There was a wave of information flow from red to grey stripes, and it seems that the larger anatomical homeostasis loop came first in terms of information flow from energy to differentiation state. Taken together, these analyses reveal the dynamics of the information flow in the tissue, we observed an emergent behavior when the "organizer" region change over time, stress is used as a communication system with the neighbours, and the higher level of homeostasis (anatomical homeostasis) came first in the dynamics suggesting that top-down information is key to manipulate the dynamics of the tissue.



\section{Experimental results on repatterning of planaria}

The sudden remodeling of cell fates after a long period of apparently quiescent, stable pattern in our model (see Figure \ref{fig:longterm}) predicts that something similar may occur in biological systems. Might some phenotypes that seem complete in fact need to be followed out for much longer to observe the true dynamic?  We indeed found this to be the case \cite{bischof2021formation}, with planarian flatworms \cite{cebria2018rebuilding} regenerating after exposure to U0126, a blocker of ERK/MAP Kinase signaling which plays an important role in many model systems of regeneration \cite{owlarn2017generic, tasaki2011erk, umesono2013molecular, yun2014sustained}. Several studies have shown that ERK inhibition immediately following amputation leads to the formation of headless animals \cite{owlarn2017generic, tasaki2011erk, umesono2013molecular}. The headless animals that regenerate after ERK inhibition had previously been assumed to have reached a terminal, stable morphology \cite{owlarn2017generic}. However, monitoring the headless animals for 18 weeks, we observed a remarkable phenotype in which some of the headless animals suddenly  began to repattern, with some regaining a wild-type single-headed morphology. 

\begin{figure}[ht!]
\centering
  \includegraphics[width=0.8\linewidth]{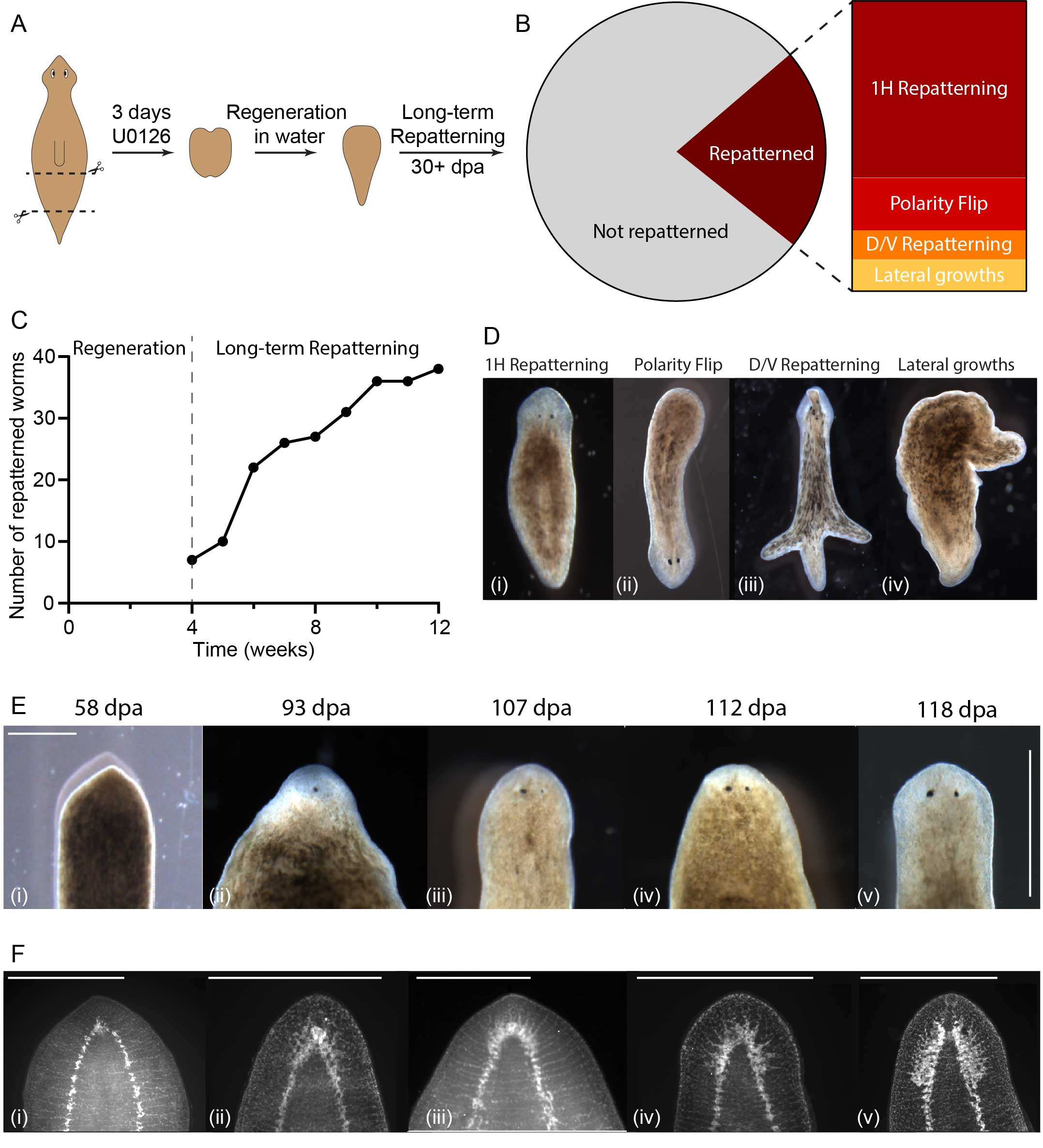}
  \caption{Headless animals can regain normal morphology over long timeframes. \textbf{A.} Illustration of the experimental procedure in which pre-tail fragments were treated for 3 days with U0126 to induce headless animals which were then observed for changes in morphology over long timeframes. \textbf{B.} Out of 181 headless worms tracked individual over time, 22\% repatterned, while the remaining 78\% did not change their morphology. The repatterned animals fell into 4 different categories – 61\% Single-head (1H) repatterning, 18\% polarity flip, 10\% D/V repatterning, and 11\% Lateral growths. \textbf{C.} Repatterning rate over time showing steady increase in the number of repatterned animals from 4 to 10 weeks after the initial amputation. \textbf{D.} Representative images of the 4 different repatterning types – i) 1H repatterning, ii) Polarity flip, iii) D/V repatterning (arrow marking dorsal outgrowth), and iv) Lateral growths. \textbf{E.} Repatterning to regain a single-headed morphology in one animal tracked over time. Scale bar 500 $\mu$m. \textbf{F.} Samples fixed and stained with synapsin antibody at different stages of repatterning, arranged into a putative neural repatterning timeline. Scale bar 500 $\mu$m.}
  \label{fig:experiment}
\end{figure}

To quantify this phenomenon, we defined repatterning as any morphology change occurring after 4 weeks post-cutting. In headless animals observed from when they were cut until their death (n=181), 22\% showed some form of sudden repatterning even though regeneration had completed and morphogenesis had ceased weeks prior (Figure \ref{fig:experiment}B). Importantly, this repatterning occurred without any intervention. The remaining 78\% of headless animals did not show any change in morphology except for shrinking in size, as headless animals are unable to feed and therefore lifetime allometrically scale down over time \cite{oviedo2003allometric}. We observed repatterning from 4 weeks onwards with the majority of repatterning occurred between 4 and 10 weeks (Figure \ref{fig:experiment}C). After 10 weeks only sporadic new repatterning was observed, although some worms began forming new heads as late as 18 weeks after cutting. This spontaneous repatterning that allows the re-establishment of a normal morphology many weeks after regeneration is completed suggests the existence of tissue processes that operate on a time scale much longer than previously known.
Anatomically, the observed repatterning fell into 4 different categories. The most common repatterning type was single-headed repatterning where a normal wild-type morphology was regained (61\% of all repatterning worms, Figure \ref{fig:experiment}B, D-i). The second most common phenotype was a polarity reversal, in which, following fissioning of a headless animal, the posterior blastema formed into a head (18\%, Figure \ref{fig:experiment}B, D-ii). The remaining two categories represent the rare animals that exhibited growths which did not lead to the formation of a head, either in the form of dorsal outgrowths (D/V repatterning, 10\%, Figure \ref{fig:experiment}D-iii) or lateral outgrowths (11\%, Figure \ref{fig:experiment}D-iv).   
When headless animals repatterned to form a new head, the pigmentation of the round anterior end began to lighten and the tissue flattened out, forming a structure resembling a blastema (Figure \ref{fig:experiment}E). Once the blastema formed, first one eyespot appeared (Figure \ref{fig:experiment}E-ii) and then the second one formed with a few days delay (Figure \ref{fig:experiment}E-iii), along with the head reshaping to regain the typical morphology (Figure \ref{fig:experiment}E-iv and v). Animals were stained using synapsin antibodies to visualize the regrowth of the underlying brain tissue at intervals during the repatterning process. This showed that early brain repatterning occurred via the formation of a cluster of neural tissue at the anterior rounding of the ventral nerve cord (Figure \ref{fig:experiment}F-i). Brain structures formed progressively from there, first in an apparently unorganized manner (Figure \ref{fig:experiment}F-ii), before expanding and reforming into a well-organized brain resembling that of a normal animal (Figure \ref{fig:experiment}F iii-v). While the repatterning process resembles normal head regeneration in its progression, the timeframe was distinctly slower than normal head regeneration \cite{cebria2002dissecting}, taking up to 25 days from the first observation of changes at the anterior to a fully formed head.

\section{Discussion}

One of the key open questions in evolutionary developmental biology, as well as basal cognition \cite{e24060819, levin2019computational}, is how single-cell capacities (competency in physiological and metabolic spaces) scale up to enable systems to solve problems in morphogenetic space (anatomical homeostasis), such as axial polarity and body-wide positional information axes. Because intelligence can be defined as competency in navigating arbitrary problem spaces, and all agents are fundamentally composites of other parts, it is important to understand how evolutionary forces and generic dynamics \cite{ goodwin1982development, goodwin2000life, newman2019inherency} work together to scale up collectives and pivot their simple homeostatic properties into more complex capabilities. Here, we presented an evolutionary simulation of the scaling of goals from cell-level metabolism to embryo-wide axial positional information. This system is a model of multi-scale homeostasis that illustrates how cells can join into collectives that solve problems in new spaces. We found that evolution was able to use the shared stress system to coordinate across space and time, and developed error-minimization and  homeostatic capacities (robustness to perturbation). The resulting tissue was found to be robust to external perturbation, using stress was both instructive for initial morphogenesis and necessary for long-term  survival. 

While  maintaining allostasis, the tissue also  tried  to get rid of the blue  cells  after  development, taking the collective temporarily away from the French flag target morphology, to then allow homeostatic mechanisms to drive it back to normal (resulting in long-term maintenance and survival of the tissue, it corresponds in the TAME framework to delayed gratification - the  ability to  not get  trapped in a local  minima). This ability to move away from the goal in order to reach a better solution later (following not only minimum-distance paths) is an important capability of some cognitive systems. Degrees of this kind of “delayed gratification” in cybernetic systems enable more sophisticated behavior, along the continuum from simple “roll downhill” mechanisms to complex problem-solving systems \cite{levin2021technological}. Interestingly, neither of these things were directly selected for during the evolutionary cycle, revealing how some capabilities of multi-scale homeostatic systems can emerge as a kind of “free lunch”, without direct selection pressure \cite{davies2022synthetic, kauffman1993origins}. 

Our analysis reveals the emergent functional connections between the single-cell and the anatomical homeostatic loops. Each cell has one goal, to survive and it corresponds to be in the appropriate state in order to receive energy (with the other members of the collective), e.g. navigating in a 1-dimensional metabolic continuous space but by discretizing it: the new emergent problem  space  solutions  has three instances (blue, red, white) corresponding to different cell fates along a positional information axis. On the other hand, the collective/tissue has a morphogenetic goal, the French flag and it is simultaneously  navigating a problem space of $3^n$ instances (n: number of cells). 

\begin{figure}[t!]
\centering
  \includegraphics[width=\linewidth]{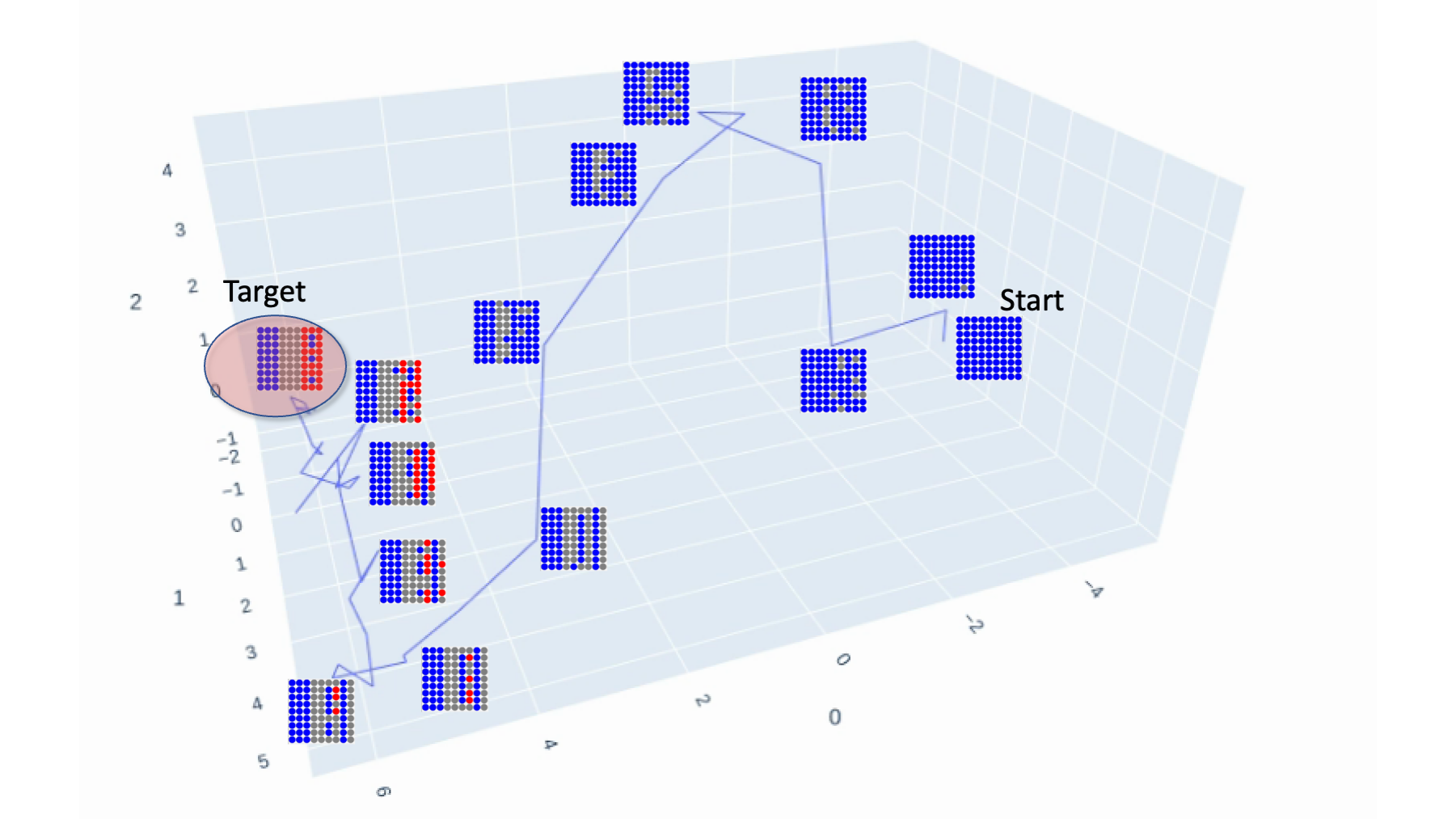}
  \caption{Visualizing the self-organization of development as navigation  in morphospace.  PCA projection  in 3D of the tissue  states  during the resolution  of the French flag (first experiment).  The target  represents the region in the morphospace  resolving  the French flag problem (during development, without any artifical interventions) enough to be compatible with survival (i.e., a primary body axis that enables high functional fitness).}
  \label{fig:nav_morphospace}
\end{figure}

The observation that planarians are capable of recovering a wild-type bodyplan by spontaneously triggering a regeneration-like process, after maintaining a stable abnormal morphology for long periods of time, poses interesting questions about how abnormal morphologies are first maintained and how they are eventually detected to trigger correction. In other organisms there is some evidence that remodeling of tissues to fit new morphologies can be part of regeneration, such as the remodeling of transplanted tail tissue into a limb in amphibia \cite{farinella1956transformation}.  The repatterning reported here is distinct from previous studies where it was shown that general injury signals can induce regeneration in headless animals \cite{owlarn2017generic}, as here there is no triggering injury of any kind. The newly demonstrated ability to mount a regenerative response without any injury to trigger it is consistent with the idea that a missing tissue response (MTR) may not be necessary for regeneration to occur \cite{tewari2018cellular} and that subtle  internal dynamics may be sufficient to initiate change long after the process appears to have reached a stable point. Our finding that the spontaneous repatterning is slower than normal regeneration is also consistent with other examples of regeneration in the absence of the MTR \cite{tewari2018cellular}.  The timepoint at which repatterning starts appears to be stochastic within the population. The observation of sudden initiation of repatterning in a seemingly quiescent tissue raises a key question: what process in the tissue serves as a “clock” that enables remodeling at a particular time frame? It must be distinct from the on-going and obvious morphogenetic processes because remodeling initiates suddenly, on a background of constant cell identity distributions.

Recently, Chou et al found a dynamic in bacterial biofilms similar to the clock and wavefront model describing  somitogenesis \cite{chou2022segmentation}. Interestingly, they observed that nitrogen stress response in the biofilm can form a complex pattern of concentric rings – revealing that already at the time of social microbes, spatial distribution of stress responses occurred. Their results show that the development of bacterial biofilm collectives is driven by  an oscillatory process that can locally amplify nitrogen stress responses, such that spores may develop in different areas of the biofilm even under a more abundant nutrient environment (not only in the starved interior of the biofilm). Interestingly, in our model, we observe a similar behavior with a subtle  oscillatory stress response of the cells in different areas of the tissue (see Figures \ref{fig:ff}B and Figure \ref{fig:ff_info_analysis}A) allowing the formation of the French flag configuration. We found that the pattern generated by the stress response in our simulated tissue defines spatial regions where cells differentiate into the different stripes, a similar mechanism we can find in the biofilm when various strains of bacteria differentiate into spores. This behavior is similar to somitogenesis where a homogenous tissue becomes segmented (transcriptionally and anatomically). In our simulation and in the biofilm, stress seems to play a key role and convey an information leading to differentiation of the cells and bacteria similarly to the clock and wavefront model. Thus, similar (conserved) dynamics could be relevant across bacterial, metazoan, and in silico minimal models of the scaling of spatial patterning from single-cell metabolic and stress dynamics.

The communication system is fundamental for collective intelligence. In our scheme, communication is mediated via the gap junctions,  a well-known system for coordinating physiological and morphogenetic activity which has also been proposed to be an essential complement to enhancing collectivity \cite{palacios2009heterotypic, mathews2017gap, levin2019computational}. In our simulation,  three types of molecules can be exchanged: the morphogen, the stress molecule and its counterpart, the "anxiolytics". In our simulation, stress plays a key role. Stress is an ambiguous and transdisciplinary concept we can find from material science to physiology and ecology \cite{parker1999stress, romero2004physiological, zimmerman2004calculation}, encompassing different meanings: certain stimuli can be stressors; the emergency responses to the stressor is defined as the stress response; and the chronic stress is  the over-stimulation of the emergency responses \cite{romero2004physiological}. In addition, the definition of these 3 concepts are mutually-dependent. A stimulus that initiates a stress response is a stressor, but the physiological or behavioral response is considered a stress response if it is initiated in response to a stressor. One solution has been to define stressors as stimuli that disrupt or threaten to disrupt homeostasis \cite{chrousos1992concepts}, but the concept of homeostasis has its own limitations \cite{schulkin2003rethinking}. A more general definition is that stressors are unpredictable and/or uncontrollable stimuli \cite{levine1991id}. Interestingly, this last definition fits with the use of the stress system in our simulation. Stress was increasing concurrently with a decrease in local active information storage in the tissue as seen in Figure \ref{fig:ff_info_analysis}B. In our scheme, evolution used the communication system as a stress system that was activated when the memory in the past was not effective for predicting the future. This relates to the active inference framework and the free-energy principle, both of which that have recently been developed to integrate homeostasis and allostasis \cite{corcoran2020allostatic}, as applied to morphogenetic systems \cite{pio2022active, kuchling2019morphogenesis}.

Network structure is key to collective intelligence \cite{chmait2016factors, schreiber2000measuring, liu2013physarum}. It has been shown that flat, fully-connected, network structures provide the most efficiency for collective to resolve a task since that type of structure maximizes the aggregation of information received from the members of the collective \cite{chmait2016factors}. However, this kind of network is very costly as it necessitates $n(n-1)$ connections and is probably not biologically realistic. Indeed, in most  cell networks, cells are only connected to their near neighbours. The stress system can be seen as another communication system and could play a role in changing the network structure to a flat (fully-connected) structure, at least for some part of the network and thus facilitating information aggregation (see \cite{mcmillen2021beyond} for a review of non-local cell communication modalities). 

One purpose of these kinds of simulations is to develop protocols that could enable robotics and machine learning approaches to benefit from the kind of robustness and problem-solving ability that is observed in biology. In AI,  the fact that all intelligences are collective intelligences has not been emphasized  \cite{levin2021technological}, but there are recent attempts for swarm robotics and deep learning to develop new methods based on collective intelligence \cite{mordvintsev2020growing, ha2021collective}. Deep learning has started to integrate the approach with adversarial networks with two networks working together \cite{goodfellow2020generative}. Artificial collective intelligence is usually studied from the point of view of swarm intelligence or human society \cite{kennedy2006swarm, mulgan2018artificial, rubenstein2014programmable} but morphogenetic systems present several characteristics lacking in swarms: a (growing) grid network architecture, the informational wiping property (cells don't know the origin of a given signal), or a different communication system closer to the information processing of a cellular automata that can only communicate with its direct neighbours. This work is one step toward integrating  the scaling of cognition findings into artificial intelligence by demonstrating dynamics that can lead to desired emergent properties across scales.

Lastly, it is interesting to consider the relevance of this system for basal cognition \cite{lyon2006biogenic, lyon2021reframing}. It has been argued that body patterning and behavioral control seems to share a common origin, not only via the mechanisms of ion channels and neurotransmitters, but also via the evolutionary pivot of solving problems from physiological to morphogenetic and ultimately 3D behavioral spaces \cite{fields2020morphological, e24060819}. Indeed, several somatic tissues exhibit evidence of learning and basal cognition, including cardiac \cite{zoghi2004cardiac}, bone \cite{turner2002bone} and pancreatic tissues \cite{goel2013learning}, in addition to the large literature on learning and decision-making in microbes and other unicellular organisms \cite{siccardi2016boolean, watson2010associative, stockwell2015yeast, prentice2016directional, barvitenko2018integration, gabalda2018recurrence}. These capacities are very old \cite{lindsay1876mind, bose1906plant} and the molecular apparatus of higher cognition (ion channels, neurotransmitters and synaptic mechanisms) was already present in our unicellular ancestors \cite{baluvska2016having}. Brains and neurons have been speed-optimized by evolution from other cell types \cite{torday2016evolution}. Somatic cells did not lose their cognitive repertoire and computational capabilities during their transitions to multicellularity and in becoming part of metazoan swarms (bodies): they scaled them to pursue larger anatomical goals \cite{pezzulo2015re}. The symmetry between development/regeneration and traditional cognition  is not  simply an  emergent result of hardwired processes, but  the emergent result of a very plastic, context-dependent system that  achieves invariant patterning outcomes under  uncertainty.  Biological systems have remarkable capacities to achieve the same morphogenetic outcomes despite a range of perturbations, such as different starting conditions, different numbers and sizes of cells, and various interventions (reviewed in \cite{pezzulo2016top, levin2022collective}). We simulated the  transitions from  single-cell homeostasis to anatomical homeostasis using  stress in this  work,  we tested the  hypothesis that  a minimal  evolutionary framework is sufficient  for small,  low-level  setpoints of metabolic homeostasis in cells  to scale up into collectives (tissues) which  solve  a problem in morphospace (see Figure \ref{fig:nav_morphospace}). These transitions take place in a continuum from body patterning to cognition, we propose that evolution pivoted the collective intelligence of cells during morphogenesis of the body into traditional behavioral intelligence by scaling up the goal states at the center of homeostatic processes. This work is a first step towards a quantitative understanding of the scaling of cognition.

\section*{Acknowledgements}

We thank  Hananel  Hazan, Benjamin Levin, and Santosh  Manicka  for helpful  discussions, and Julia Poirier for assistance with the manuscript. LPL thanks Antoine and Isabelle Lopez and Flagey 10 for the general support. We gratefully acknowledge the support of Grant 62212 from the John Templeton Foundation. The opinions expressed in this publication are those of the author(s) and do not necessarily reflect the views of the John Templeton Foundation.

\newpage
\bibliographystyle{plain}
\bibliography{ref.bib}

\end{document}